\documentclass[twocolumn]{aastex701}
\usepackage{rotating}
\usepackage[normalem]{ulem}
\usepackage{xcolor}
\usepackage{amsmath}

\newcommand{\mr}{\mathrm}

\received{July 14, 2025}
\submitjournal{ApJ}

\begin{document}

\title{Accretion from a Shock-Inflated Companion: Spinning Down Neutron Stars to Hour-Long Periods}

\correspondingauthor{Savannah Cary}
\email{scary@berkeley.edu}

\author[0000-0003-1860-1632]{Savannah Cary}
\affiliation{Department of Astronomy, University of California, Berkeley, CA 94720-3411, USA}
\email{scary@berkeley.edu}
\author[0000-0002-1568-7461]{Wenbin Lu}
\affiliation{Department of Astronomy, University of California, Berkeley, CA 94720-3411, USA}
\affiliation{Theoretical Astrophysics Center, University of California, Berkeley, CA 94720-3411, USA}
\email{wenbinlu@berkeley.edu }
\author[0000-0002-4209-7408]{Calvin Leung}
\affiliation{Department of Astronomy, University of California, Berkeley, CA 94720-3411, USA}
\email{calvin_leung@berkeley.edu}
\author[0000-0001-9195-7390]{Tin Long Sunny Wong}
\affiliation{Department of Physics, University of California, Santa Barbara, CA 93106, USA}
\email{twong31@physics.ucsb.edu}

\begin{abstract}

Recent observations have unveiled a population of pulsars with spin periods of a few minutes to hours that lie beyond the traditional ``death line.'' If they originate from neutron stars (NSs), the existence of such ultra-long period pulsars (ULPs) challenges our current understanding of NS evolution and emission.
In this work, we propose a new channel for disk formation based on NSs born in close binaries with main-sequence companion stars. Using a hydrodynamic simulation of supernova-companion interactions, we show that a newborn NS may gravitationally capture gas as it moves through the complex density field shaped by the explosion. For a binary separation of $20\rm~R_\odot$ and a companion mass of $4\rm~M_\odot$, we find the occurrence fraction for disk formation around unbound NSs to be $\sim10\%$.
By modeling the disk evolution and its interaction with the NS, we find a bimodal distribution in spin periods: canonical pulsars with $P\lesssim10\rm\,s$ are the ones who lack disks or whose magnetospheres never interacted with the disk, and ULPs with $10^3\lesssim P<10^5\rm\,s$ are produced when the system undergoes a short-lived ``propeller'' phase during which the NS undergoes rapid spin-down. Such ULPs are formed under strong initial dipolar magnetic field strengths $B_0\gtrsim10^{14}\rm\,G$, with a formation rate of $10^{-4}\rm\,yr^{-1}$ in the Milky Way. We also find that a small population of pulsars with moderate magnetic field strengths ($10^{13}\lesssim~B_0\lesssim10^{14}\rm\,G$) and relatively slow initial periods ($P_0\gtrsim0.1\rm\,s$) evolve to $P\sim10^2\rm\,s$, filling the gap between the bimodal distribution. Thus, our model provides a unified explanation for pulsars beyond the ``death line.''
\end{abstract}


\section{Introduction}
\label{sec:intro}
Prior to the past few years, radio pulsars as neutron stars (NSs) were known to have periods of $\sim$~0.002-12~s \citep[][\url{http://www.atnf.csiro.au/research/pulsar/psrcat}]{Kaspi2017, pulsarcat}. The lack of periods $>$12~s has been  historically referred to as the ``death line'': a traditional pulsar that spins down due to magnetic field decay would stop producing radio emission at these long periods, as it would no longer be able to produce electron-positron pairs along the open field lines anchored on the polar caps \footnote{Observational biases in traditional pulsar searching also play some role in the lack of long period pulsars \citep[e.g.,][]{Lazarus2015,vanHeerden2017}.}\citep[e.g.,][]{Chen1993, Zhang2000}. 

The first few peculiar pulsars that pushed this upper limit included PSR J0250+5854 with a period of 23.5~s \citep{tan2018}, PSR J1903+0433 with a period of 14~s \citep{han2021}, and PSR J0901-4046 with a period of 76~s \citep{caleb2022}. Driven by these discoveries, the death line and various aspects of the long-term evolution of NSs have been reevaluated, including magnetic field decay, inclination angle evolution, and possible influences by a fall-back accretion disk \citep[e.g.,][]{Igoshev_Popov2018, Kou2019, Ronchi2022,Gencali2023,zhou2024}. 

More recently, the story has become increasingly difficult to interpret with the discovery of even slower radio pulsators whose periods do not immediately suggest NSs. These discoveries include CHIME J0630+25 \citep{chime421}, GLEAM-X J162759.5-523504.3 \citep[GLEAM-J1627,][]{18min}, GPM 1839–10 \citep[GMP-1839,][]{21min}, ASKAP J1935+2148 \citep[ASKAP-J1935,][]{54min}, and ASKAP/DART J1832-0911 \citep[ASKAP/DART-J1832,][]{Li2024, Wang2024}, with periods of 421~s, 18.18~min, 21~min, 54~min, and 44~min, respectively. Table~\ref{table:ULPs} lists these observed pulsators and their properties. Most of these sources lack precise localizations for multi-wavelength counterpart searches. An exception is ASKAP/DART-J1832, whose spatial coincidence with the supernova remnant\footnote{\citet{Wang2024} cautiously noted the non-negligible probability of chance coincidence.} G22.7-0.2 and radio polarization properties suggest a NS origin \citep{Li2024}; this is also supported by the pulsed X-ray emission during an outburst with phase-averaged luminosities of $L_{\rm X}\sim10^{33}\rm\, erg\,s^{-1}$ \citep{Wang2024}. 

Another motivation for the NS nature of some of these long-period sources is the magnetar candidate IE 161348-5055 (IE-1613), who is embedded in a young supernova remnant (SNR) RCW103 and has a rotational period of $6.7$~hr \citep{DeLuca2006}. Although IE-1613 has not been detected in the radio so far, other magnetars show transient radio emission likely powered by magnetic energy rather than rotational energy \citep[e.g.,][]{camilo06_radio_emission_magnetar, levin10_radio_loud_magnetar, israel01_magnetar_radio, zhu23_SGR1935_radio_pulses}.

However, two other long-period radio pulsators, GLEAM-X J0704‑37 with a 2.9~hr period and ILT J1101+5521 with a 2.1~hr period \citep[][]{HurleyWalker2024_MDWARF,ruiter2024}, have been detected with optical counterparts and are more likely to originate from binary systems with an M-dwarf (MD) orbiting around a white dwarf (WD). Indeed, spectroscopic measurements of GLEAM-X J0704‑37 by \citet{rodriguez25_orbital_solution_3hr_source} confirmed the MD-WD binary nature of this source and showed that the period of radio pulsations corresponds to the orbital period. In this scenario, the radio emission is likely powered by magnetospheric interactions between a strongly magnetized WD and the MD companion \citep{Katz2022, LoebMaoz2022, Qu25_WDMD_magnetospheric_interactions}. This is analogous to the systems of AR Scorpii and J1912–4410 \citep{Marsh2016,marcote17_AR_Scorpii_compact_radio, Pelisoli2023}, although in these two well-studied systems the WD's rotation is not synchronous with the orbit (likely due to accretion spin-up episodes).

Though this may suggest a WD-MD origin for these recently discovered long-period pulsators, many of the sources without optical counterparts (to deep limits) seem to be isolated. To date, no isolated magnetic WD has been observed to show radio pulsations. In fact, \citet{Beniamini2023} argued that \textit{isolated} magnetic WDs are inconsistent with observations of these long-period sources. On one hand, WDs have very low magnetic dipole spin-down powers, $L_{\rm sd}$. If we take a typical WD radius of $R_{\rm WD}\sim 5\times10^3\rm\, km$, then the spin-down power is $L_{\rm sd}\sim 5\times10^{27}\mr{\,erg\,s^{-1}} (B/10^9\mr{\,G})^2 (P/\mr{hr})^{-4}$ for surface magnetic field strength $B$ and spin period $P$. In the dipole spin-down picture, the peak luminosities of the observed radio pulses ($L\gtrsim 10^{31}\rm\, erg\,s^{-1}$) would generally require magnetic fields that are much stronger ($B\gg 10^9\rm\, G$) than the most strongly magnetized WDs known \citep{chime421, 54min}. On the other hand, powering the radio emission by magnetic field decay is also problematic, because the magnetic energy budget $E_{\rm B}\sim 10^{43} (B/10^9\mr{\,G})^2\rm\, erg$, combined with the requirement of a cooling age of $\tau \gtrsim 10^8\rm\, yr$ (as constrained by the optical upper limits), would give a rather low luminosity $L_{\rm B}\sim E_{B}/\tau\lesssim 3\times10^{27}\mr{\,erg\,s^{-1}} (B/10^9\mr{\,G})^2 (\tau/10^8\mr{\,yr})^{-1}$.
Therefore, if these sources are isolated, observations favor the picture of a strongly magnetized NS or magnetar where the radio pulses are magnetically powered (instead of rotationally powered).

For the purpose of this paper, we will call the sources showing radio pulsations with periods $P\gtrsim 10^3\rm\, seconds$ \textit{ultra long-period pulsars} (ULPs). We assume that a fraction of ULPs are isolated NSs, despite that no ULP (besides IE-1613) has yet been associated with a NS so far. Since NSs are generally born rapidly spinning  \citep[e.g.,][]{Popov2012, Noutsos2013}, we focus on how to slow down an isolated NS's spin to these ultra long periods. 

A widely studied mechanism is the braking torque from an accretion disk --- for strong NS magnetic fields and low accretion rates, the interactions between the accretion flow and the NS magnetosphere may slow down the spin of the NS \citep{davidson73_NS_accretion}. Spin down from accretion disks has indeed been observed in the high-mass X-ray binary system 4U 1954+319, which contains a NS with a spin period of 5.4 hrs and a red supergiant companion star of mass $\gtrsim 7 ~\rm M_\odot$ \citep{enoto14_slowNS_in_HMXB, hinkle20_slowNS_in_HMXB}. In fact, the companion star may go supernova and be kicked from the system, leaving behind an isolated, slow-spinning NS \citep{mao2025binaryoriginultralongperiod}.

There is also observational evidence of NSs with disk accretion in their evolutionary history. One such example is pulsar PSR1257+12, who has two Earth-like planets that orbit it \citep{Wolszczan1992}. Additionally, the magnetars 4U 0142+61 and 1E 2259+586 show near-infrared (NIR) emission that is likely from a debris disk heated by the NS's X-ray emission \citep{perna00_disk_around_NS, Hulleman2004,Wang2006, Ertan2007,Kaplan2009}, although \citet{Hare2024} argued for a magnetospheric origin based on the large amplitude variability of the NIR/optial emission. Furthermore, the magnetar candidate IE 1613 (with a 6.7 hr spin period) went into an outburst phase with a NIR counterpart that may be produced by an X-ray heated debris disk \citep{tendulkar17_NIR_counterpart, esposito19_NIR_counterpart}.

Previous studies on disk-magnetosphere interactions have shown that a NS can be slowed down to minute to hour long periods
by an accretion disk of initial mass $M_{\rm d}\sim 10^{-6} - 10^{-3} \ \textrm{M}_{\odot}$ \cite[with initial magnetic field strengths in the range $\sim 10^{12} - 10^{15}$~G, e.g.,][]{Gencali2023, Ronchi2022, fan2024, Tong2023}. In this framework, \citet{DeLuca2006} and \citet{Tong2016} showed that IE-1613's spin period of 6.7~hrs could be achieved with an initial disk mass of $\sim 10^{-5} \ \textrm{M}_{\odot}$ and initial magnetic field strength of $\sim 10^{15}$~G. \citet{HoAndersson2017} further emphasized the need for IE-1613 to have a long-lived ($\gtrsim10^3\rm\, yr$) and spatially extended disk (radius $\gg R_\odot$) with very low accretion rates of the order $\sim 10^{-12} \ \textrm{M}_{\odot}\, \mr{yr}^{-1}$.

Most prior studies rely on supernova fallback to form the initial disk around the NS. Numerical simulations indeed yield sufficient fallback masses from $10^{-4}~ \rm M_\odot$ to a few $0.1~\textrm{M}_{\odot}$ \citep[e.g.,][]{Michel1988, Lin1991,Ugliano2012, Perna2014, ertl20_fallback, Janka2022}. \citet{Janka2022} found net specific angular momenta of the order $10^{16} $ to $ 10^{17} \textrm{cm}^2 \ \textrm{s}^{-1}$, corresponding to very compact disks with circularization radii of the order $10^6 $ to $10^8~\textrm{cm}$. It should be noted that the net angular momentum in the simulations summarized by \citet{Janka2022} comes from the asymmetric supernova mass ejection but not stellar rotation --- the latter has been considered by \citet{Perna2014}. The general conclusion is that the fallback disks are initially very compact with extremely high peak accretion rates of $10^{-6}$ to $10^{-2} ~ \rm M_\odot\rm\,s^{-1}$ on a timescale of $10$ to $10^2\rm\, s$ after the core collapse. It is theoretically unclear if these disks will survive to the very late evolutionary phases of $\gtrsim 10^3\rm\, yr$ in order to sufficiently spin down a NS. In fact, \citet{Perna2014} argued that these super-Eddington accretion disks may be quickly depleted by very strong radiation driven winds, so they likely do not survive on kyr timescales.

With this said, it is unclear whether the fallback disk model is sufficient in spinning down isolated NSs to the periods of recently observed pulsators. In this paper, we propose an alternative way to create a gas disk around a NS. Our model is based on a binary system where one of the stars explodes in a supernova and the newly created NS can gravitationally capture gas from the shock-impacted, close-by companion star\footnote{This is not the first time it has been conceptualized that an isolated NS can form an accretion disk from post-supernova-binary-companion interaction. In fact, \citet{PhinneyHansen1993} suggested that a NS could be kicked towards its companion, causing it to tidally disrupt and form a disk.  Although our model slightly differs, it goes to show that the authors already thought carefully about how a binary-originating NS can become an isolated NS with a disk.}. Our model predicts much more extended disks than the fallback disks considered in previous works, with disk radii evolving from the initial values of the order $10^7$--$10^{11}\rm\, cm$ to $\gtrsim 1\rm\,AU$ on kyr timescales. 

Our model is schematically described in \S\ref{sec:ourmodel}.
Then, \S\ref{sec:hydro} describes our hydrodynamic simulations of a supernova in a binary system. In \S\ref{sec:masscapture}, we calculate the mass and angular momentum of the gas captured by the NS by post-processing of the simulation, and in \S\ref{sec:diskevo} we describe the long-term evolution of the accretion disk formed around the NS. Then, in \S\ref{sec:periods}, we consider the evolution of the NS spin due to the disk-magnetosphere interactions and present the results on final spin period distribution. \S\ref{sec:discussion} discusses these results in the context of ULP observations, and in \S\ref{sec:conclusion} we summarize our results.

\begin{table*}
\label{table:ULPs}
\centering
\begin{tabular}{|c|c c c c c c|} 
\hline 
\textbf{Name} & $P$ & $b$ & DM  & Counterparts &  Ref. Paper & Source?\\
 & (s) & ($\circ$) & ($\textrm{pc} \ \textrm{cm}^{-3}$) & & & \\
\hline
PSR J0250+585 & 23.5 & -0.5 & 45 & None &  \citet{tan2018} & NS\\
PSR J0901-4046 & 76 & 3.7 & 52 & None &   \citet{caleb2022} & NS \\
AR Scorpii & 118 & 19 &  (117~pc) & M-Dwarf & \citet{Marsh2016} & WD\\J1912–441 & 318 & -22 & (238~pc) &  M-Dwarf & \citet{Pelisoli2023} &WD \\
CHIME J0630+25 & 421 &  7.1 & 23  & None  & \citet{chime421} & ?\\
GLEAM-X J1627 & 1091 & -2.6 & 57 & None  & \citet{18min} & ?\\
GMP-1839 & 1318 & -2.1 & 274 & None &   \citet{21min}& ?\\
ASKAP/DART-J1832 & 2656 & 0.1 & 465--480 & X-ray  & \citet{Li2024} & ?\\
ASKAP-J193 & 3225 & 0.7 & 146 & NIR &   \citet{54min} & ?\\
GLEAM-X J0704-37 & 10440 & 13 & 37 & M-Dwarf &  \citet{HurleyWalker2024_MDWARF} &WD \\
ILT J1101+5521 & 7350 & 56 & 16 & M-Dwarf &  \citet{ruiter2024} &WD \\
IE-1613 & 24000 & -0.4 & -- & X-Ray &  \citet{DeLuca2006} &NS \\
\hline
\end{tabular}
\caption{List of known radio pulsators. From left to right we include their emission period, galactic latitude, dispersion measure, known observed counterparts, the reference paper, and the compact object thought to be responsible for the emission. All are observed in the radio except IE-1613, which is an X-Ray source.}
\end{table*}

\section{Our Mechanism} \label{sec:ourmodel}

We consider NSs formed from supernovae in binary systems. The vast majority of massive main-sequence stars ($M\gtrsim 10~\rm M_\odot$) relevant for core-collapse supernovae are found in binary, triple, or higher-order systems \citep{Raghavan2010, Sana2012, Duchene2013, offner23_stellar_multiple}. The evolution of these massive stars is shaped by various binary mass-transfer effects that could result in stripped-envelope supernovae; there is strong evidence that the progenitor stars for Type Ib/c supernovae lost their Hydrogen envelope via mass transfer with a binary companion \citep[e.g.,][]{Sana2012,Langer2012,Yoon2012,Smartt2015,Prentice2018,Smith2011}. An important (but theoretically not fully understood) outcome is that about $30\%$ of core-collapse supernovae are from these stripped-envelope stars \citep[e.g.,][]{Li:2011aa, Drout2011, Eldridge:2013aa, Lyman2016, Taddia2018, woosley21_SNIbc}. For this reason, we choose to focus on type-Ib/c systems for our paper. 

In a binary system, the supernova ejecta will interact with the companion star, which is most likely a main-sequence star. It has been shown through hydrodynamic simulations of core-collapse supernovae that the companion star will be shock heated due to the ejecta’s impact and expand to 5-100 times its original radius \citep{Hirai2014,Hirai2018,Ogata2021,Chen2023}. Additionally, assuming that the core-collapse supernova results in a NS, the NS will get a large velocity kick of a few $100\rm\, km\,s^{-1}$ \citep{Hobbs2005}. In most cases, this kick will unbind the NS from its companion, and on its path, it will encounter its companion's shock inflated envelope. As it passes through the envelope, it will undergo Bondi capture \citep{Bondi1952} (see Figure~\ref{fig:athena_ex} for the evolution of such a system). The expanded envelope will have a density gradient that may provide enough net angular momentum for Bondi capture compared to the supernova fallback model (see Figure~\ref{fig:bondi_cap}).

\citet{Hirai2018} studied a system with an initial companion star of mass 10~$\textrm{M}_{\odot}$ and radius 5~$\textrm{R}_{\odot}$, 
and separation distances of 20-60 $\textrm{R}_{\odot}$. From largest to shortest separation distances, their companion star inflated to be $30 \sim 200 \textrm{R}_{\odot}$ a short time after explosion.
\citet{Chen2023} looked at models of separation distance 3-8 $\textrm{R}_{\odot}$ and masses of 3-8 $\textrm{M}_{\odot}$. They also showed that the companion can grow to be about a magnitude larger than its original size after the shock impact. \citet{Ogata2021} found similar results for the same separation distances, with masses of 3 and 15 $\textrm{M}_{\odot}$, and ejecta mass of 2 $\textrm{M}_{\odot}$. \citet{Liu2015} studied type-Ib/c supernovae with lower ejecta mass of 1.4~$\textrm{M}_{\odot}$ impacting on less massive companion stars of 0.9 and 3.5 $\textrm{M}_{\odot}$, and they found that the inflation of the shock-impacted companion star strongly depends on the orbital separation.

Since we expect a fraction of the stripped-envelope supernovae to have periods of a few days prior to explosion \citep{Podsiadlowski1992,Sana2012, Sana2013, Gagliano2022}, in this paper we adopt a separation distance of 20 $\textrm{R}_{\odot}$ prior to explosion. Furthermore, we will use an ejecta mass of $5 \ \textrm{M}_{\odot}$, which is consistent with observations of stripped-envelope supernova \citep{Barbarino2021, Taddia2015, Drout2011, Taddia2018, Lyman2016,Prentice2019}. We consider a main-sequence companion star of mass $4 \ \textrm{M}_{\odot}$. We expect our  final NS spin period distribution to depend strongly on the orbital separation (and weakly on the companion star mass and ejecta mass), but we leave a detailed exploration of the larger parameter space to a future work. In this paper, we focus on the effects of different NS kick velocities and hence only consider one orbital configuration. 
Our system
has three phases. First, the heating of the companion star by the supernova blast wave is simulated using the hydrodynamic set-up described in \S\ref{sec:hydro}. Then, the formation of the fallback disk is modeled numerically by post-processing the simulation outputs in \S\ref{sec:masscapture}. Finally, the disk-NS interaction, including the angular momentum transfer between the two, is analytically modeled to late times (up to $10^6$ yrs) in \S\ref{sec:diskevo} and \S\ref{sec:periods}.

\section{Hydrodynamic Simulation} \label{sec:hydro}

\subsection{Setup}


We use the hydrodynamic code \textsc{athena++} \citep{athena} to simulate the ejecta-companion interaction, with our code adopted from \citet{Wong2024} and \citet{Bauer2019}. We solve the hydrodynamical equations on a Cartesian grid with a fiducial $512^3$ resolution in the co-moving frame of the companion star. A lower $256^3$ resolution simulation is performed and the small difference between these two runs justifies the the convergence of our results (see Appendix~\ref{sec:resolution}). We also run the $256^3$ resolution simulation for a much longer time to confirm the late-time evolution of the shock-heated star. We adopt an ideal gas equation of state with $\gamma = 5/3$, and utilize the multigrid capabilities of \textsc{athena++} for self-gravity \citep{TomidaStone2023}. Although the supernova ejecta's pressure is dominated by radiation, here we focus on the response of the companion star. Since most of shock-heated regions are gas pressure-dominated, a choice of $\gamma=5/3$ provides a reasonable approximation for our problem.

The companion star is taken to be an $n=3$ polytrope with mass $M_\mathrm{C} = 4 \ \textrm{M}_{\odot}$ and radius $R_{\mathrm{C}} / \textrm{R}_{\odot} \approx 
\left( M_{\mathrm{C}} / \textrm{M}_{\odot} \right)^{0.6}.$ The NS progenitor's mass is the sum of its ejecta and the mass of the NS remnant, $M_{\mathrm{ej}} + M_{\mathrm{NS}}$. For this paper, we assume a NS mass of $M_{\rm NS} = 1.4 \ \textrm{M}_{\odot}$. With an ejecta mass of $M_{\rm ej} = 5 \ \textrm{M}_{\odot}$, this gives our pre-explosion star a mass of $6.4 \ \textrm{M}_{\odot}$. 

The supernova explosion 
originates from a distance $a_{sep} = 20~\textrm{R}_{\odot}$ from the companion star's center, and the ejecta freely expands until hitting the companion star. In \textsc{athena++}, we initialize the outer edge of the ejecta to be $5 \ \textrm{R}_{\odot}$ from the center of explosion. The ejecta has a maximum velocity $v_{max}$ of $2 \times 10^4 \ \textrm{km} \ \textrm{s}^{-1}$ with a total explosion energy $E = 10^{51}$~erg. The density of the ejecta is taken to be a broken power law as described in \citet{Chevalier1989} and \citet{Kasen2010}, with profiles $\rho_i \propto r^{-\delta}$ and $\rho_o \propto r^{-n}$ for the inner and outer regions, respectively. We take $\delta$ and $n$ to be 1 and 10, respectively. Assuming an initial orbit in the x-y plane, we give each ejecta fluid element a y-velocity offset to account for the orbital velocities of the two stars prior to explosion. For stars of mass $6.4 \textrm{M}_{\odot}$ and $4 \textrm{M}_{\odot}$, we find that the relative velocity between the two stars is $314.9~\textrm{km} \ \textrm{s}^{-1}$.

To model adiabatic cooling, similar to \citet{Bauer2019}, the internal energy of a fluid element at a given time is set to be a fraction of the kinetic energy, dictated by its radius from explosion center $r$ at time $t$. The kinetic energy density is such that $KE = 1/2 \rho v^2$, where $v$ is the velocity and $\rho$ is the density of the fluid element. The pressure at a given radius is $P = \left( \gamma - 1 \right) \cdot U$, where $U$ is our internal energy density. For adiabatic expansion, the ratio between internal and kinetic energy is $P/(\rho v^2)\propto \rho^{\gamma-1}\propto r^{-3(\gamma-1)}\propto t^{-3(\gamma-1)}$ (for $v=$constant). Realistically, the SN ejecta is radiation dominated, so we expect $\gamma = 4/3$ and hence $-3(\gamma-1) = -1$. Using this scaling, we set the internal energy density as $U = KE \cdot r_{equil} / r$, where $r_{equil}$ is the equilibrium radius (at which $KE = U$). Our $r_{equil}$ is $0.08 a_{sep}$, such that as the ejecta reaches the companion its internal energy is $\sim$ 8\% of the kinetic energy.

Lastly, like \citet{Wong2024}, we use the passive-scalar capability in \textsc{athena++} \citep{athena}, where a passive scalar is a tracer that does not modify fluid properties. Two passive scalars are added, one for the ejecta and one for the companion. This will allow us to distinguish between ejecta and stellar mass captured by the NS later on. 

Table~\ref{tabel:properties} summarizes our simulation input parameters. 

\begin{table}
    \centering
    \caption{Simulation Parameters}
    \label{table:sims}
    \begin{tabular}{l|l l}
        \hline
        \textbf{Parameter} & \textbf{Label} & \textbf{Value}  \\ 
        \hline
        Grid Size & -- & $512^3$ \\
        Domain & -- & $x,y,z \in \left[-20,20 \right] \textrm{R}_{\odot} $ \\
        Companion Mass & $M_{\rm C}$ & $4~\textrm{M}_{\odot}$  \\
        Ejecta Mass & $M_{\rm ej}$ &  $5~\textrm{M}_{\odot}$  \\
        Ejecta Energy & $E$ & $10^{51}~\textrm{ergs}$  \\
        Max Ejecta Velocity & $v_{max}$ & $2\times 10^4~\textrm{km}~\textrm{s}^{-1}$ \\
        Orbital Velocity & $v_{orb}$ & 314.9 $\textrm{km}~\textrm{s}^{-1}$ \\
        Separation Distance & $a_{sep}$ &  20  $~\textrm{R}_{\odot}$ \\
    \end{tabular}
    \label{tabel:properties}
\end{table}

\subsection{Hydrodynamic Results}

\begin{figure*}
    \centering
    \includegraphics[width=\textwidth]{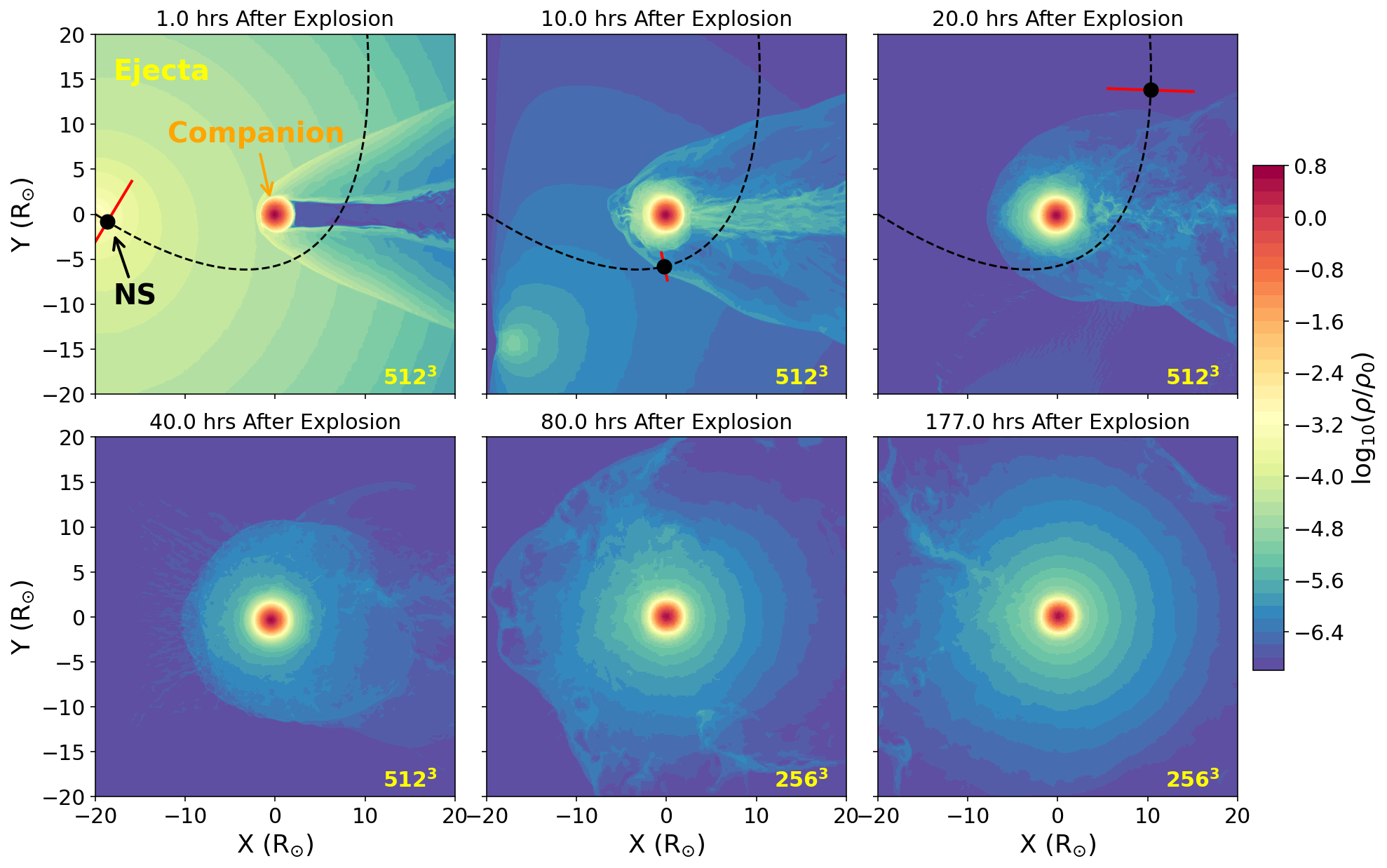}
    \caption{\textbf{Density slices in the $z=0$ plane.} Overlaid is an example NS trajectory (black dashed line), with a black circle representing the current NS position. In this example, the NS trajectory is inside the x-y plane with a kick velocity $300~\textrm{km} \ \textrm{s}^{-1}$. The red bar represents the critical Bondi radius $r_{\rm B}$ for gas capture (see Equation~\ref{eqn:rb}), dependent on the relative velocity between the NS and the local gas, as well as the local sound speed of the gas. Color indicates gas density in units of $\rho_0 = \textrm{M}_{\odot}~\textrm{R}_{\odot}^{-3}$. The top and bottom left panels are for the fiducial resolution of $512^3$ grid (used in this paper). We only ran the $512^3$ simulations for $\sim 40$~hrs, as an unbound NS would not interact with the companion star past this time. The last two panels are for a separate simulation with a lower-resolution $256^3$ grid, and is only used to confirm the late-time evolution of the companion. }
    \label{fig:athena_ex}
\end{figure*}

Figure~\ref{fig:athena_ex} shows our simulation results at different timesteps. Note that at $t>40$~hrs, we show a lower resolution simulation on a grid of size $256^3$ using our previously described parameters. The NS is expected to have a kick on the order of a few hundred $\textrm{km} \ \textrm{s}^{-1}$ \citep{Hobbs2005}, and will no longer interact with the companion envelope by these late times ($>40$~hrs). However, the lower-resolution simulation is ran for a longer duration to confirm the secular expansion of the companion. 

As expected, the ejecta interacts with the companion star and the shock-inflation begins within a day of the initial interaction. The companion star eventually relaxes into a spherically symmetric profile with an envelope that expands to several times its original radius by $\sim$1 week. The purpose of this paper is to discuss accretion disk formation around an unbound NS and its subsequent evolution into a ULP, so we will not discuss the companion star long-term evolution on timescales $\gg 1\rm\, day$. 

Additionally plotted in Figure~\ref{fig:athena_ex} is an example NS kick trajectory. The NS will pass through and capture gas surrounding the companion star. Note that we do not include the NS gravitational potential in the simulation. Instead, in the following section we post-process the saved data from the simulation to study the gas capture by the NS.

\section{Mass Capture} \label{sec:masscapture}

\subsection{NS trajectory and Bondi capture}
 \begin{figure}
     \centering
     \includegraphics[width=0.45\textwidth]{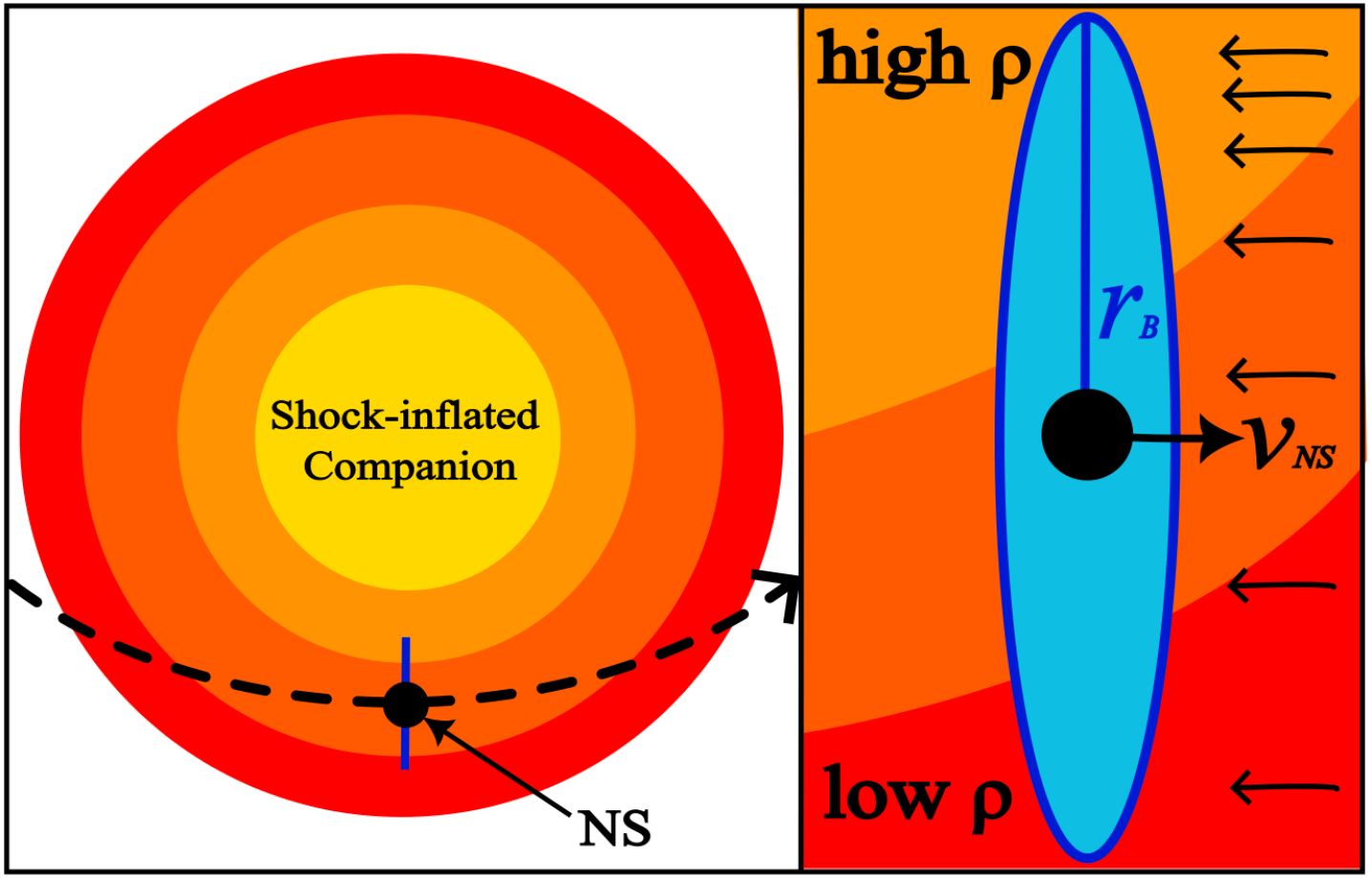}
     \caption{\textbf{Schematic for gas capture.} The panel on the left shows a NS being kicked through its companion's shock inflated envelope. The NS will travel through the expanded envelope of the companion and undergo gas capture described by the Bondi radius $r_B$ \citep{Bondi1952}, represented with a blue line. We zoom in on the NS and show the plane for Bondi capture in the right panel. Because the envelope will be more dense towards the companion star's center. In this scenario, there is a density gradient in the captured gas, which  provides sufficient angular momentum to circularize into a disk.}
     \label{fig:bondi_cap}
 \end{figure}

 \begin{figure}
     \centering
     \includegraphics[width=0.45\textwidth]{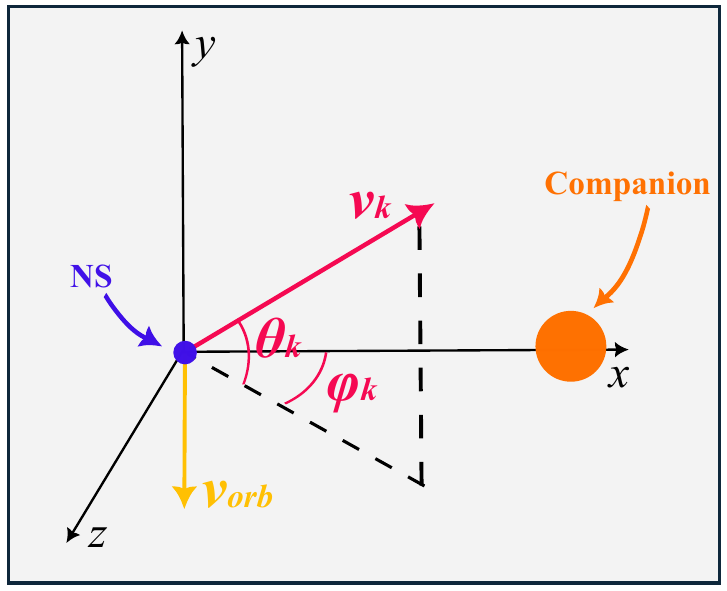}
     \caption{\textbf{NS Kick Schematic.} The companion star is situated along the x-axis. The NS kick velocity is described by magnitude $v_k$, the latitude angle $\theta_k = \pi/2 - \theta$ away from the x-z plane (where $\theta$ is the polar angle from the y-axis), and the azimuthal angle $\varphi_k$. In yellow is the orbital velocity vector prior to explosion, $\vec{v}_{\rm orb} = -v_{\rm orb}~ \hat{y}$. The final initial velocity vector is given by $\vec{v}_{\rm NS} = \vec{v}_{\rm orb} + \vec{v}_k$.}
     \label{fig:angles}
 \end{figure}

In this section, we consider that a NS is launched from the center of the explosion with original orbital velocity $\vec{v}_{\rm orb} = v_{\rm orb} ~ \hat{y}$ plus an additional kick velocity $\vec{v}_k$. We test 3 different kick amplitudes: 300, 400, and $500~\textrm{km} \ \textrm{s}^{-1}$. 

We save 3D snapshots of the simulation, and solve for the NS's kick trajectory within each snapshot. We assume the NS is only influenced by the companion's gravitational force. The NS's kick trajectory is based on kick amplitude $v_k$ and kick angles $\theta_k$ and $\varphi_k$ as described by Figure~\ref{fig:angles}. Here, $\theta_k = \pi/2 - \theta$ is defined as the latitude angle away from the $x$-$z$ plane, where $\theta$ is the polar angle from the y-axis. 
The azimuthal angle $\varphi_k$ is defined as the angle between the projected velocity in the $x$-$z$ plane and the $x$-axis. Our final probability density functions are based on the full parameter space of kick angles $\theta_k \in \left[-\pi/2, \pi/2 \right]$ and $\varphi_k \in \left[0, 2\pi \right]$. However, we only consider the kick angles for which the NS becomes unbound from the companion star and ends up isolated (satisfying the constraints on the observed ULPs).
We reduce the computation by considering the fact that the entire system is symmetric with respect to the equatorial ($x$-$y$) plane; a kick with azimuthal angle $\varphi_k$ will have the same mass capture as another trajectory with azimuthal angle of $-\varphi_k$. We also assume that no ejecta is captured for kicks away from the companion star ($\pi/2 <\varphi_k < 3\pi/2$), as in those cases, the NS directly moves away from the companion star. Thus, we practically only compute the mass captures for the unbound trajectories for a dense grid of $-\pi/2 < \theta_k < \pi/2$ and $-\pi /2 < \varphi_k < \pi/2$.

As the NS travels, it will gravitationally capture gas that is sufficiently close to its path, as demonstrated in Figure~\ref{fig:bondi_cap}. For mass capture, we adopt a cross section perpendicular to the NS's instantaneous velocity vector with an area of $\sigma = \pi r_B^2$, where $r_{\rm B}$ is the Bondi radius \citep{Bondi1944} approximately given by 
\begin{equation}
    r_B \approx \frac{2GM_{NS}}{v_{rel}^2 + c_s^2}.
    \label{eqn:rb}
\end{equation}
Here, $G$ is the gravitational constant, $c_s$ is the sound speed of the fluid element, and $\vec{v}_{rel} = \vec{v}_{NS}-\vec{v}_{fl}$ is the relative velocity between the NS ($\vec{v}_{NS}$) and the fluid ($\vec{v}_{fl}$).

Given the NS position and velocity vector, we find which fluid elements undergo Bondi capture at a given time. The rates of mass and angular momentum capture, $\dot{M}$ and $\dot{\vec{L}}$, at a given time are given by
\begin{equation}
    \dot{M}
    \approx \int_0^{r_{\rm B}} d r'\int_0^{2\pi} d\varphi' \rho ~ r' ~ v_{rel},
    \label{eqn:Mdot}
\end{equation}
and
\begin{equation}
    \dot{\vec{L}}
    \approx \int_0^{r_{\rm B}} d r'\int_0^{2\pi} d\varphi' \rho \, r'\, v_{rel} \left( \vec{r}^{\,\prime} \times \vec{v}_{rel} \right),
    \label{eqn:Ldot}
\end{equation}
where $\rho$ is the density of a given fluid element at position $\vec{r}^{\,\prime}$ in the NS's frame. At each time, we consider polar coordinates $\vec{r}^{\,\prime} = (r', \phi')$ in the plane perpendicular to the NS's velocity vector ($\vec{v}_{\rm NS}$). We interpolate the density, pressure, and velocity fields of the fluid elements to obtain their values on a 2D grid of $(r', \phi')$ in the perpendicular plane, and then sum up the contributions to the mass and angular momentum captures by those fluid elements that satisfy $r' < r_{\rm B}(r', \varphi')$. Since the fluid velocity $v_{fl}$ and sound speed $c_s$ are typically smaller than the NS velocity, the above approximations are reasonable. Lastly, we exclude any fluid elements that are more gravitationally bound to the companion star than the NS ($M_C/r_C >M_{NS}/r_{NS}$, where $r_C$ is the distance from the companion star center and $r_{NS}$ is the distance from the NS center).

We dump 3D snapshots of the simulation on a time interval of $\delta t = 0.5 \times t_{scale}$, where $t_{scale} = (\textrm{R}_{\odot}^3/G\textrm{M}_{\odot})^{1/2} = 1593\rm\, sec$ is the time unit used in our simulation. We solve for the position of the NS within each snapshot, and calculate $\dot{M}$ and $\dot{\vec{L}}$ using Equation~\ref{eqn:Mdot} and Equation~\ref{eqn:Ldot}. This results in one data point per snapshot. More information on time resolution testing can be found in Appendix~\ref{sec:resolution}.

An important issue with our approach is that a fraction of the slower moving supernova ejecta is inevitably captured by the NS; this is conceptually similar to the case of fallback accretion. However, due to our simplistic set-up for the spherically symmetric ejecta profile and our ignorance of the radiation pressure (as well as heating sources such as $^{56}$Ni), the amount of fallback accretion obtained in our simulation cannot be trusted --- a more realistic simulation of the supernova explosion is required to study the realistic fallback accretion. For this reason, we use the following procedure to remove the contribution to the mass and angular momentum capture of the NS from fallback accretion.

\begin{figure}
    \centering
    \includegraphics[width=0.47\textwidth]{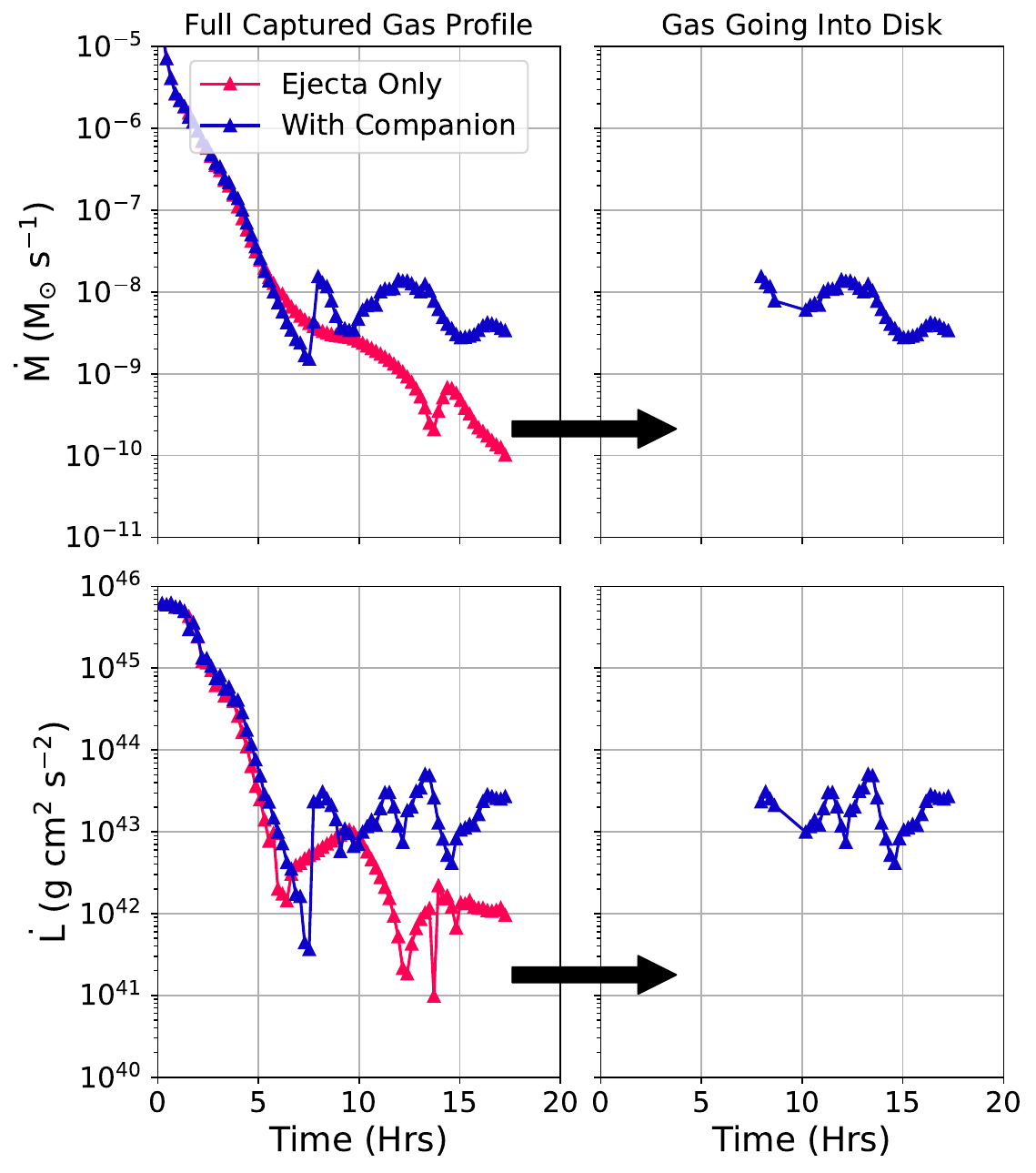}
    \caption{\textbf{Process of removing fallback accretion.} Plotted in blue are the mass- and angular momentum-capture rates using Equation~\ref{eqn:Mdot} and Equation~\ref{eqn:Ldot} from our simulations. In red we plot the same for our companion-less simulation (where we allow the ejecta to freely expand). Each data point represents the mass-capture rate per snapshot (every $0.5 \times t_{\rm scale} = 0.5 \times \sqrt{R_\odot^3/G M_\odot} = 796.5\rm\, sec$). Data points that differ by a factor of 2 between the two simulations are considered to be ``true'' mass and angular momentum captured by the NS. The dip in the companion-less simulation results around $\sim13$ hours is due to part of the Bondi radius extending outside the simulation domain as the NS passes close to the simulation boundary.}
    \label{fig:removed_ej}
\end{figure}

To distinguish what is simple free-flowing ejecta, we run a separate simulation similar to the one described in \S~\ref{sec:hydro}, except we do not include the companion star and instead allow the ejecta to expand freely. We compare the mass-capture rates at each time between the simulations with and without a companion star. We remove the data points where the mass-capture rates between the two simulations are within a factor of two of each other. Since most of the fallback accretion occurs in the first few hours before the NS reaches close to the companion star, we find that our choice of the threshold for the data removal successfully differentiates between early-time fallback accretion and the late-time mass capture of gas that has been strongly affected by the companion star. 

Figure~\ref{fig:removed_ej} shows an example of mass-capture and angular momentum-capture rates for a given kick velocity as specified by $v_k$, $\theta_k$, and $\varphi_k$. The precise amount removed depends on the NS kick velocity, but we generally find that the removed mass is on the order of $10^{-2} \ \textrm{M}_{\odot}$ and that the removed angular momentum is of the order $10^{48}$~erg~s. If this ejecta were to form an accretion disk, it would result in an extremely compact disk with radius $\sim 10^6$~cm. 

After this filtering process, the total captured mass $M_{d,0}$ is $\sum_{n} \dot{M}_n \delta t$, where $n$ is the number of data points that survive our fallback-accretion removal procedure, and $\delta t = 0.5 \times t_{scale} $ is the time resolution of our Bondi-capture calculation. Similarly, $L_{d,0,x}=\sum_{n} \dot{L}_{x,n} \delta t$, $L_{d,0,y}=\sum_{n} \dot{L}_{y,n} \delta t$, and $L_{d,0,z}=\sum_{n} \dot{L}_{z,n} \delta t$. 

With our initial disk mass $M_{d,0}$ and angular momentum $L_{d,0} = |\vec{L}_{d,0}|$, the initial radius of the disk $R_{d,0}$ is obtained under the assumption of a circular Keplerian disk
\begin{equation}
   L_{d,0} = M_{d,0}  \sqrt{G M_{NS} R_{d,0}} . \label{eqn:R}
\end{equation}
Therefore, we obtain the total captured mass $M_{d,0}$ and initial circularization radius $R_{d,0}$ for each realization of kick velocity. This allows us to study the subsequent evolution of the accretion disk later in the next section.

Lastly, we use our simulation's passive scalars to track how much final captured mass comes from the initial ejecta and how much comes from the companion star. Although we evolve our disks using the total captured mass (albeit, with the fallback accretion removed), we are also interested in seeing if a disk and subsequent NS spin-down can be created from the companion's stellar mass alone. We call these two different types of disk models ``total mass disks'' and ``stellar mass disks.''

\subsection{Mass Capture Results}

\begin{figure*}
    \centering
    \includegraphics[width=\textwidth]{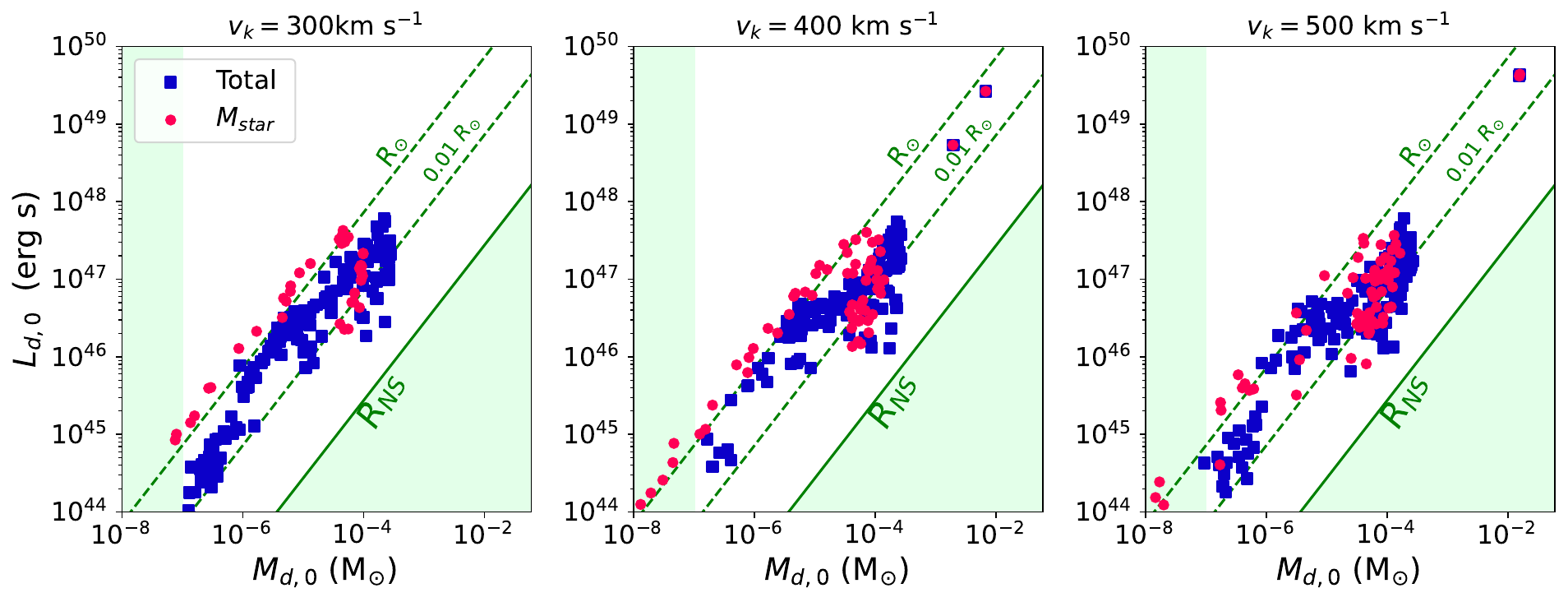}
  \caption{\textbf{Captured disk masses and angular momenta.} Lines of equal radii are plotted using Equation~\ref{eqn:R}. The green shaded regions represent disks that have been disqualified due to our radius cutoff of $10^6$~cm and mass cutoff of $10^{-7}~\textrm{M}_{\odot}$. Blue squares are of the total disk mass, whereas red circles are disks created considering only companion stellar mass.}
    \label{fig:MLR}
\end{figure*}

Using our three kick velocities, we calculate the fraction of kick angles that result in unbound and bound NSs. For both the sake of this paper and simplicity, if the NS comes within the initial radius of the companion star, we call this a tidal disruption event (TDE) and remove it from these fractions. We calculate the fraction of unbound NSs to be 0.79, 0.87, and 0.95 for kick amplitudes of 300, 400, and 500~$\textrm{km} \ \textrm{s}^{-1}$, respectively. The fraction of bound NSs is 0.19, 0.12, and 0.05 for kick amplitudes of 300, 400, and 500~$\textrm{km} \ \textrm{s}^{-1}$, respectively. Meanwhile a fraction of 0.01, 0.01, and 0.001 will result in a TDE for kick amplitudes of 300, 400, and 500~$\textrm{km} \ \textrm{s}^{-1}$, respectively.

Not all kick angle realizations will result in a disk that we evolve in \S~\ref{sec:diskevo}. First, in this work, we restrict ourselves to unbound orbits that would lead to an isolated NS, as suggested by ULP observations. 
Furthermore, to limit uncertainties in the mass capture, we ignore disks that have masses less than $10^{-7} \ \textrm{M}_{\odot}$ (such low mass disks have little effect on the NS spin anyway). 
As shown in Figure~\ref{fig:MLR}, these limits can disqualify a small fraction of cases. Each data point in Figure~\ref{fig:MLR} represents a different kick angle that are sampled according to a grid of isotropic distribution of kick directions. The disks in our model have masses of $10^{-7} \sim 10^{-2}~\textrm{M}_{\odot}$, and angular momentum of $10^{44} \sim 10^{49}~\textrm{erg~s}$.

\begin{figure}
    \centering
    \includegraphics[width=0.47\textwidth]{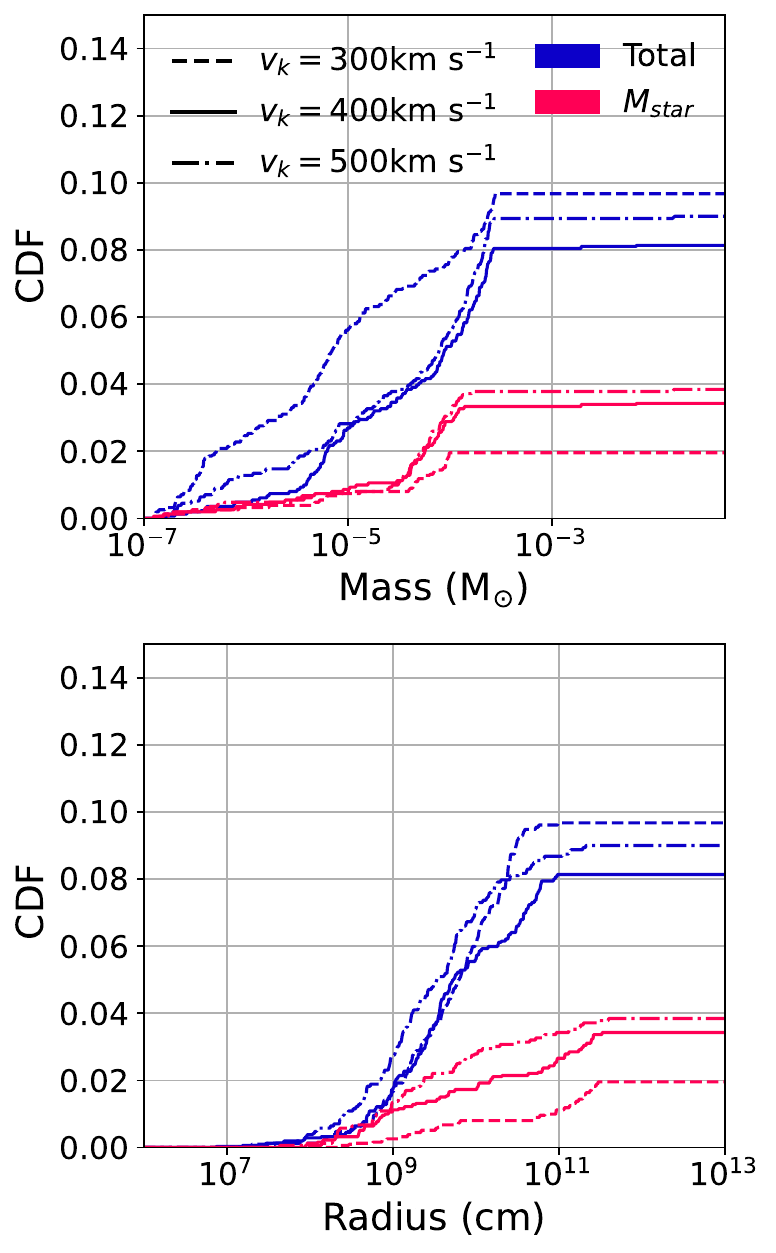}
    \caption{\textbf{Fraction of disks formed} for three kick velocities: 300, 400, and 500~$\textrm{km} \ \textrm{s}^{-1}$. Out of all possible kicks ($\theta_k = [- \pi/2, \pi/2]$, $\varphi_k = [0, 2\pi]$), we calculate the fraction of disks formed in unbound orbits that have mass greater than $10^{-7} \ \textrm{M}_{\odot}$, and circularization radius greater than  $10^{6} \ \textrm{cm}$. Blue represents the disks we calculate in \S\ref{sec:masscapture}, whereas red  represents the disks created considering only companion stellar mass.}
    \label{fig:frac}
\end{figure}

Figure~\ref{fig:frac} shows the fraction of kick realizations that result in accretion disks, for three different kick velocities. This fraction is taken out of all possible kick angles, where we only count accretion disks that will form around isolated (unbound) NSs.  With our limits imposed, we expect to create a disk around an isolated NS 8-10\% of the time for our systems. If instead we only consider stellar mass disks, this percentage drops to 2-4\%. 

Although not always true, much of the contribution of mass and angular momentum of a given disk are from the supernova ejecta fluid elements whose trajectories have been strongly affected by the companion star due to shocks. The NS flies past the companion star on timescales much shorter than the time it takes for the shock-heated outer envelope of the companion star to expand significantly. Thus, the contribution from the stellar mass to the disk is usually small. 

In general, lower velocity kicks (300~$\rm km \ {s}^{-1}$) are more likely to form disks, as a higher kick velocity leads to a lower critical radius for Bondi capture (Equation~\ref{eqn:rb}). However, we also find that NSs with an initial kick magnitude of 400~$\rm km \ {s}^{-1}$ form disks less frequently than 500~$\rm km \ {s}^{-1}$. This is likely because an initial kick velocity of 500~$\rm km \ {s}^{-1}$ has a higher probability of resulting in unbound orbits, yet for this higher velocity the companion envelope has not expanded enough for the NS to undergo sufficient gas capture (so the fraction does not exceed the 300~$\rm km \ {s}^{-1}$ kick realizations).

\section{Disk Evolution} \label{sec:diskevo}

After acquiring our disk masses and circularization radii, we analytically solve for the evolution of each disk, which viscously spread and accrete onto the NS. Our goal is to solve for the time evolution of the disk mass $M_d(t)$, disk radius $R_{\rm d}(t)$, and mass accretion rate $\dot{M}_{\rm d}(t)$. We adopt the simplest one-zone model \citep{Shen2014} which captures most of the physics governing the long-term evolution of the disk. The interactions between the disk and the NS magnetosphere are considered in a postprocessing manner, as we expect the effects of accretion feedback on the evolution of the outer disk to be negligible. We assume that the disk rotates near the Keplerian speed and is in hydrostatic equilibrium in the vertical direction.

We find that each of our disks have different mixes of contributions from the ejecta and companion stellar mass, resulting in varying types of chemical compositions. Throughout this section, we assume a solar abundance; however, we note that the disk composition is not exactly solar because of the captured Hydrogen-deficient ejecta. 

At any given time $t$, the mid-plane sound speed $c_s$, scale-height $H$, Keplerian angular velocity $\Omega_k$, and the surface density $\Sigma$ of the disk are given by
\begin{align}
   c_s^2 = \frac{P}{\rho} \mbox{,}\quad & H = \frac{c_s}{\Omega_k} \mbox{,}\quad \\
   \Omega_k = \left(\frac{G M_{NS}}{R_d^3}\right)^{1/2} \mbox{,}\quad & \frac{M_d}{\pi R_d^2}=\Sigma = 2\rho H , 
\end{align}
with pressure $P$ and density $\rho$. The shear viscosity is given by the \citet{Shakura1973} $\alpha$-prescription: $$\nu_{vis} = \alpha c_s H.$$ Motivated by the observations of dwarf novae, previous works have considered a variable $\alpha$ model \citep{Hameury1998, Ichikawa1992, Vishniac1996}. We follow the method in \citet{Hameury1998} by adopting
\begin{multline} \label{eqn:alpha}
    \textrm{log}\alpha =  \textrm{log}\alpha_{\rm cold} \\ + \left( \textrm{log}\alpha_{\rm hot} - \textrm{log}\alpha_{\rm cold} \right)\times \left[ 1+ \frac{2.5\times10^4 ~\textrm{K}}{T}\right]^{-1},
\end{multline}
where $\alpha_{\rm hot}$ = 0.1 represents a disk in the hot regime, and $\alpha_{\rm cold}$ = 0.02 represents the cold regime. $T$ is the mid-plane temperature of the outer disk near radius $R_d$. We further note that $\alpha$ changes with gas composition in the disk  \citep{2018ApJ...857...52C}, however, we show in Appendix~\ref{sec:alpha} that our final results do not depend strongly on the $\alpha$ prescription we use. 

The pressure $P$ comes from radiation pressure and gas pressure such that $P = P_{gas} + P_{rad}$. With Boltzmann constant $k_{B}$, radiation constant $a$, mean molecular weight $\mu$ (fixed to be $0.6$), and proton mass $m_p$, the pressure in the disk mid-plane is:
\begin{equation}
   P =  \frac{\rho k_{B}T}{\mu m_p} + \frac{a T^4}{3}. 
\end{equation}

We solve for the mid-plane temperature $T$ from the equation of thermal equilibrium, with viscous $(q_{\rm vis}^+)$ and irradiative $(q_{\rm irr}^+)$ heating balancing radiative $(q_{\rm rad}^-)$ and advective ($q_{\rm adv}^{-}$) cooling, i.e.,
\begin{equation}
    q_{\rm vis}^{+} + q_{\rm irr}^+ = q_{\rm adv}^{-} + q_{\rm rad}^{-}.
\end{equation}
We follow \citet{Shen2014} for the expressions for $q_{\rm vis}^+, q_{\rm rad}^-,$ and $q_{\rm adv}^{-}$, and add irradiation effects from NS X-ray emission \citep[e.g.,][]{Ertan2009, Alpar2013}. We assume a constant X-ray luminosity of $L_{x} = 10^{35}\rm\, erg\,s^{-1}$, which is reasonable among known magnetars\footnote{See the magnetar catalog: \url{http://www.physics.mcgill.ca/~pulsar/magnetar/main.html}}. The irradiative heating rate is taken to be
\begin{equation}\label{eqn:irradiation}
    q_{\rm irr}^+ = \frac{1}{5}\frac{L_x}{4\pi R_d^2}\frac{H}{R_d},
\end{equation}
which is derived from the disk's flared geometry in Appendix~\ref{sec:diskgeometry}.

Putting everything together, the equation of thermal equilibrium is given by
\begin{gather} \label{eqn:heating}
\frac{9}{4} \nu_{vis}\Sigma\Omega_k^2 + \frac{1}{5}\frac{L_x}{4\pi R_d^2}\frac{H}{R_d}= \frac{M_d/t_{\rm vis}}{2\pi R_d^2}\frac{P}{\rho} + \frac{4acT^4}{3\kappa \Sigma},
\end{gather}
where $c$ is the speed of light and $\kappa$ is the temperature-dependent Rosseland-mean opacity. The opacity $\kappa$ is taken from the OPAL project \citep{IglesiasRogers1996} for temperatures $T \gtrsim 10^4$~K and \citet{Ferguson2005} for $T \lesssim 10^4$~K (including the effects of molecules and dust grains). We assume a solar composition.

We solve the above equation for the mid-plane temperature $T$, which then tells us the vertical scale-height $H$ and hence the viscous timescale,
\begin{gather}
    t_{vis} = \frac{\alpha}{\Omega}\left(\frac{H}{R_d}\right)^{-2}.
\end{gather}
The time evolution of the system is governed by the viscous timescale.

Finally, the equations of mass and angular momentum conservations are given by
\begin{gather}
    \dot{M}_d = -\frac{M_d}{t_{vis}} + \dot{M}_{\rm cap}(t),
\end{gather}
and
\begin{equation}
    \dot{L}_d = \dot{L}_{\rm cap}(t),
\end{equation}
where the mass capture rate during disk formation is given by $\dot{M}_{\rm cap}(t) = M_{d,0}/t_0$ for $t < t_0$ and $\dot{M}_{\rm cap}(t) = 0$ for $t > t_0$. The angular momentum capture rate is given by $L_{\rm cap}(t) = \dot{M}_{\rm cap}(t) \sqrt{GM_{NS} R_{d,0}}$. For simplicity, we fix $t_0 = 20\rm\, hrs$ to be the duration for the initial mass capture phase for all kick angles. We expect our results to not depend strongly on our choice of $t_0$ because the spin evolution of the NS only depends on the late-time ($t\gg 1\rm\, yr$) evolution of the disk. The angular momentum conservation equation, combined with the assumption of Keplerian rotation, gives the time evolution of the outer disk radius
\begin{gather}
    \frac{\dot{R}_d}{2R_d} = \frac{1}{t_{vis}} + \frac{\dot{M}_{\rm cap}}{M_d} \left({\sqrt{R_{d,0}/R_d} - 1}\right).
\end{gather}

\begin{figure}
    \centering
    \includegraphics[width=0.47\textwidth]{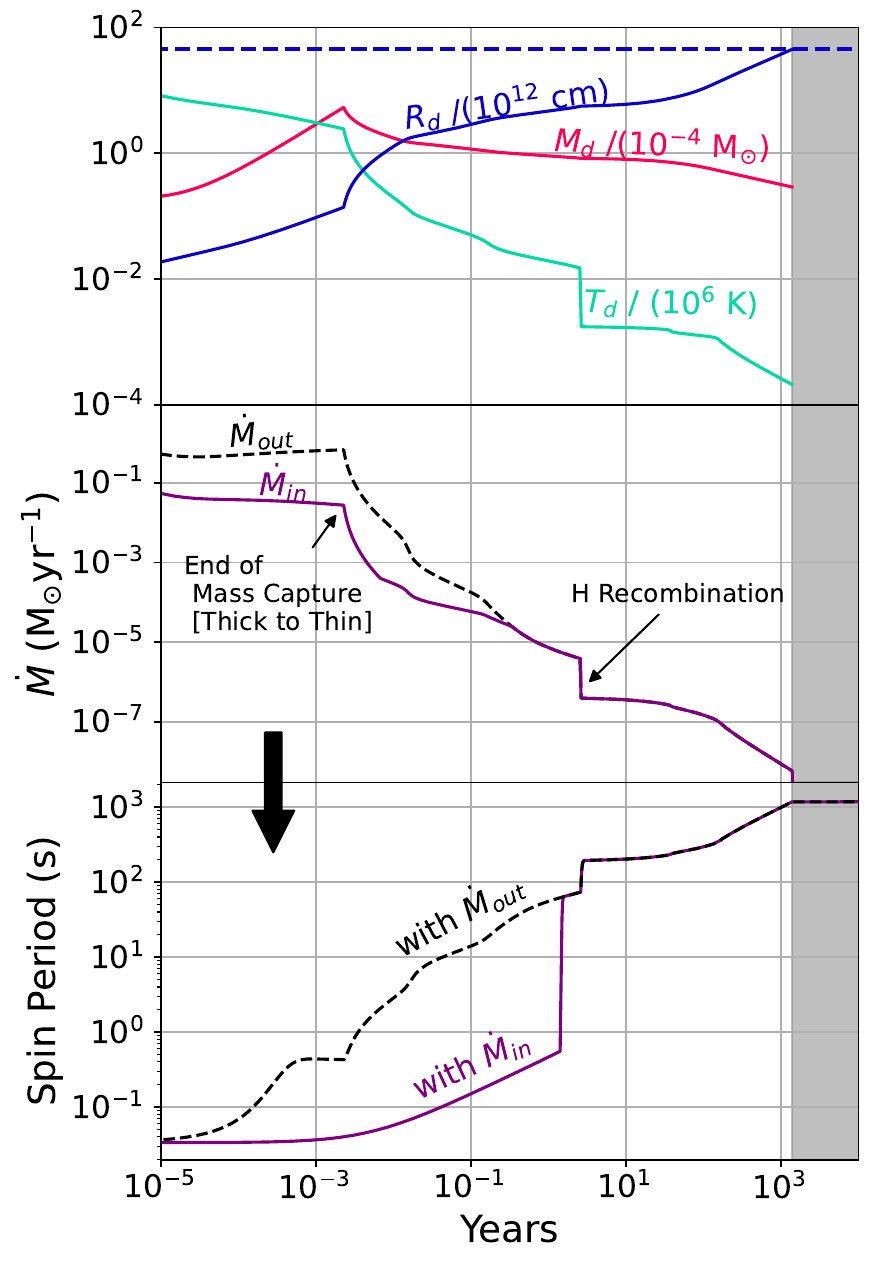}
    \caption{\textbf{Temperature, mass, radius, and accretion rate evolution of one of our disks.} The dashed horizontal line in the top panel represents the irradiation radius, in which we stop disk evolution and consider the disk to be evaporated. We include the accretion rate in the outer disk $\dot{M}_{out}$ and that in the inner disk $\dot{M}_{in}$ near the Alfv\'en radius (taking into account mass loss due to super-Eddington wind, see \S\ref{sec:periods}). The accretion rate suddenly drops in two locations: where the disk goes from thick to thin regime as mass is no longer being added to the disk, and when the opacity becomes dominated by hydrogen recombination. We further plot the spin period evolution, using our $\dot{M}_{in}$ values and methods described in \S\ref{sec:periods}. 
    }
    \label{fig:diskevo}
\end{figure}

We track the evolution of each disk realization with the captured masses and circularization radii from Figure~\ref{fig:MLR}. The evolution of one such disk is shown in upper and middle panels of Figure~\ref{fig:diskevo}, where we see that the disk undergoes two important state transitions. First, the disk goes from a radiation pressure-dominated (geometrically thick) regime to a gas pressure-dominated (geometrically thin) regime on a timescale of $t_0 = 20$~hrs as mass is no longer being added to the system. Second, once the gas in the mid-plane of the disk cools below $\sim10^4\rm\, K$, the opacity rapidly drops due to hydrogen recombination \citep[see][for a similar evolution]{yang24_disk_evolution} and the disk further collapses into an even thinner one. We note that if we instead assumed a Hydrogen-deficient gas composition, Helium recombination would set in at a higher temperature, and the disk would collapse at an earlier time. This may change the onset of the ``propeller phase'' (when the NS rapidly spins down, see \S \ref{sec:periods}), but the NS long-term evolution would follow the equilibrium spin period described in \S \ref{sec:periods}, and we do not expect our final spin results to be substantially affected. 

When the outer disk viscously evolves to even larger radii (a few to 10 AU), it is expected that the disk will experience photo-evaporation due to the NS's X-ray emission \citep{Owen2012}. The details of the photo-evaporation process are complex as the mass-loss rate due to evaporation depends on the surface temperature $T_{\rm s}$ of the disk under X-ray heating \citep{1983ApJ...271...70B}. When the sound speed of the photo-heated surface layer reaches a significant fraction of the local escape speed, we expect the gas to escape the system as a result of vertical expansion and subsequent heating due to X-ray heating. Thus, the disk evaporates when its outer radius reaches a fraction of the following characteristic radius
\begin{equation}
    r_{\rm evap} = \frac{2\mu m_{\rm p}GM_{\rm NS}}{k_{\rm B} T_{\rm s}} = 6.0\mathrm{\,AU}\, \left(\frac{T_{\rm s}}{3\times 10^4\rm\,K}\right)^{-1},
\end{equation}
where $T_{\rm s}$ is the temperature of the evaporating surface layer that is heated due to direct X-ray irradiation. In our model, we account for the effect of photo-evaporation by imposing a maximum disk radius $R_{\rm d,max} = 3\rm\, AU$. This means that we truncate the disk evolution once $R_d > R_{\rm d,max}$ and we no longer consider its effects on the NS's spin. By the time the outer disk reaches such a large radius, the mid-plane temperature has cooled down to a typical temperature of $\lesssim 100$~K and is believed to become passive anyway \citep[e.g.,][]{inutsuka05_passive_disk_100K}. We find that the disk truncation due to photo-evaporation generally occurs much later than the propeller phase during which the NS rapidly spins down (see \S \ref{sec:periods}), so our results of ULP formation are robust against the details of photo-evaporation.

\section{Neutron Star Spin}\label{sec:periods}

\subsection{Spin Evolution}\label{sec:periods_evo}

Once the disk evolution is obtained, we then consider the interactions between the disk and the NS magnetosphere.
We first define the light cylinder radius $r_{lc}$, the corotational radius $r_{co}$, and the Alfv\'en radius $r_{A}$ \citep{GhoshLamb1978} as
\begin{gather}
r_{lc} = {c/\Omega} \\
r_{co} = \left( GM/\Omega^2\right)^{1/3} \\
r_{A} = \left( \frac{3B^2R^6}{2\dot{M}_{in} \sqrt{G M}}\right)^{2/7}, \label{eqn:r_A}
\end{gather}
where $\Omega$ is the spin angular frequency of the NS, and $\dot{M}_{\rm in}$ is the accretion rate of the inner edge of the disk at the Alfv\'en radius.

At early times when the accretion rates are super-Eddington, mass removal by disk winds causes $\dot{M}_{\rm in}$ to be different from the accretion rate of the outer disk, which is given by
\begin{equation}\label{eq:Mdot_out}
    \dot{M}_{out} = M_d/t_{vis}.
\end{equation}
The characteristic radius for winds driven by super-Eddington radiation is the spherization radius \citep{Shakura1973}
\begin{equation}
r_{sph} = G M_{NS} \dot{M}_{out} / L_{Edd},    
\end{equation}
where $L_{Edd}=M_{NS} G \pi  m_p c/ \sigma_T$ is the Eddington luminosity of the system. The disk is geometrically thick at radii $r < r_{sph}$ and a strong wind is expected, causing the true accretion rate to drop towards smaller radii as a power-law $\dot{M}\propto r^p$ \citep{Blandford1999}, where $p$ is typically 0.3 - 1 according to numerical simulations \citep{Yuan2012}. In this paper, we take $p = 0.5$ as motivated by more recent simulations \citep{guo24_p_index, cho24_p_index, 2024MNRAS.532.4826T}.

If $r_A > r_{sph}$, the effects of disk wind are negligible so we take $\dot{M}_{in} = \dot{M}_{out}$ for a steady-state disk. However, if $r_A < r_{sph}$, the mass loss due to wind reduces the accretion rate near the Alfv{\'e}n radius such that
\begin{align} \label{eqn:wind}
\dot{M}_{in} = \dot{M}_{out} \left( \frac{r_A}{\min(r_{sph}, R_d)} \right)^{p}.
\end{align}
Using the accretion rate in the inner disk in Equation~\ref{eqn:r_A}, we obtain the modified Alfv{\'e}n radius
\begin{equation}
    r_A = \left( \frac{3B^2 R^6}{2 \dot{M}_{out} \sqrt{G M_{NS}}}\right)^\frac{1}{p+7/2} \left[ \min(r_{\rm sph}, R_d)\right]^\frac{p}{p+7/2}.
\end{equation}
The torques on the NS depend on the order of these three radii $r_{lc}$, $r_{co}$, and $r_A$.

If $r_A$ is greater than $r_{lc}$, then the NS will not interact with the disk and will act as an isolated pulsar. The disk begins to interact with the NS's magnetic field if $r_A \leq r_{lc}$, which causes a torque on the NS from the disk $\dot{\Omega}_{disk}$ \citep{davidson73_NS_accretion, GhoshLamb1978}. If the inner disk is moving faster than the rotational velocity of the NS ($r_{co} > r_{A}$), then the NS will experience a positive, spin-up torque. On the other hand, if $r_{co} < r_{A}$, then the NS will experience a negative torque that acts to slow the NS down; this is referred to as the ``propeller phase.'' 

Whether or not the disk interacts with the magnetic field, there will also be a spin-down torque on the NS due to magnetic dipole emission alone, denoted as $\dot{\Omega}_{mag}$. The magnitude of this torque depends on whether $r_{A}$ is greater than or less than $r_{lc}$. When the disk crosses into the magnetosphere, some field lines become open \citep{Parfrey2016,Metzger2018}, which exert an additional torque. 

At a given time, the time evolution of the NS's spin angular frequency, $\dot{\Omega} = \dot{\Omega}_{disk} + \dot{\Omega}_{mag}$, is summarized by the following equations:
\begin{equation}
    r_A > r_{lc}:\  \dot{\Omega}_{disk}  = 0,\  \dot{\Omega}_{mag} = -\frac{B^2R^6\Omega^3}{Ic^3},
\end{equation}
\begin{equation}
\begin{split}
    r_A \leq r_{lc}: \ \dot{\Omega}_{disk} &= \frac{\dot{M}_{in} \sqrt{ G M_{NS} r_A}}{I} \left( 1-\left( r_A/r_{co}    \right)^{3/2}  \right), \\
    \dot{\Omega}_{mag} &= -\frac{B^2R^6\Omega^3}{Ic^3} \left(r_{lc}/r_{A}\right)^2, 
\end{split}
\end{equation}
where whether the disk is spun up or down is encrypted in the $1-\left( r_A/r_{co}    \right)^{3/2}$ term \citep[see also][]{Eksi2005,Piro2011}. 

When the system is in the propeller regime ($r_{\rm lc} > r_{\rm A} > r_{\rm co}$), there are two spin-down torques, one from the disk ($\dot{\Omega}_{\rm disk}$) and the other one from the magnetic fields ($\dot{\Omega}_{\rm mag}$). The ratio between the two spin-down torques is given by $$\frac{|\dot{\Omega}_{\rm disk}|}{\dot{\Omega}_{\rm mag}} = \frac{3}{2} \left( \frac{r_{A}} {r_{g}}\right)^{1/2},$$ where $r_{\rm g} = GM_{\rm NS}/c^2$ is the gravitational radius of the NS. In our model, since $r_{A}\gg r_{\rm g}$, the disk torque always dominates over the magnetic torque during the propeller phase. When the disk interacts with the magnetosphere for a sufficiently long time, the NS spin comes to an equilibrium rate $\Omega_{eq}$ given by $r_{co} = r_{A}$ (ignoring $\dot{\Omega}_{\rm mag}$ in the propeller phase):
\begin{equation}\label{eqn:omega_eq}
\begin{split}
        \Omega_{eq} &= \left( G M_{NS} \right)^{5/7} \left( \frac{2 \dot{M}_{in}}{3 B^2 R^6} \right)^{3/7}\\
        &= 4.8\times10^{-2}\mathrm{\,rad\,s^{-1}}\, \left(\frac{\dot{M}_{\rm in}}{10^{-8}\,M_\odot\mathrm{\,yr^{-1}}}\right)^{3/7} B_{14}^{-6/7}.
\end{split}
\end{equation}
The scalings of $\Omega_{\rm eq}\propto \dot{M}_{\rm in}^{3/7} B^{-6/7}$ means that, when the system reaches the equilibrium state, the slowest spins are achieved at low accretion rates and strong B-fields. In fact, it is believed that NSs in high-mass X-ray binaries reach very long spin periods (up to many hours) as a result of the propeller effect due to a negative $\dot{\Omega}_{\rm disk}$ \citep[e.g.,][]{stella86_slow_NS_in_HMXB}. Turning the argument in Equation~\ref{eqn:omega_eq} around, we can solve for the critical accretion rate corresponding to the equilibrium state at a given spin period
\begin{equation}
\begin{split}
    \dot{M}_{\rm cr} &= \frac{3B^2R^6 \Omega^{7/3}}{2 \left(GM\right)^{5/3}}\\
    &= 5.7\times10^{-11}\,M_\odot\mathrm{\,yr^{-1}}\, B_{14}^2 \left(\frac{P}{10^3 \,\rm s}\right)^{-7/3},
\end{split}
\end{equation}

where $B_{14} = B/10^{14}~\textrm{G}$. We note that other works have different torque models for $\Omega_{disk}$. In the literature, the term $1-\left( r_A/r_{co}   \right)^{3/2} $ often becomes $1-\omega^{n} $, where $\omega = \left(r_A/r_{co}\right)^{3/2}$ is referred to as the ``fastness'' parameter. We take $n$ to be 1 in our calculations, however, works such as \citet{Gencali2022}, \citet{Gencali2023} took $n=2$. Our final period distribution does not depend on the choice of $n$, which only affects how rapidly the spin rate evolves towards $\Omega_{\rm eq}$ (Equation~\ref{eqn:omega_eq}) when the system enters the propeller phase. This is because we always find that the NS quickly reach $\Omega_{\rm eq}$ soon after the propeller phase (on a timescale much shorter than the disk evolution timescale, see Figure~\ref{fig:periodevo}).

Lastly, our model includes the effect of magnetic field strength decay over time. There are multiple causes for field decay, including Ohmic decay, the Hall effect, and ambipolar diffusion \citep[e.g.,][]{Goldreich1992, Pons2007}. As we consider NSs with $B \geq 10^{13}$~G, it is expected that the magnetic field decay will be dominated by the ambipolar diffusion of the solenoidal component on a timescale of \citep{Goldreich1992}
\begin{equation}
    \tau_{ambip} = 3 \times 10^5 ~ \textrm{yr}^{-1}\frac{L_5^2 T_{c,8}^2}{B^2_{14}}, 
\end{equation}
where $L_5 = L/10^5~\textrm{km}$ is the length scale for magnetic field variations (we take $L_5=1$), $B_{14} = B/10^{14}~\textrm{G}$ is our magnetic field strength at a given time, and $T_{c,8} = T_c/10^8~\textrm{K}$ is the core temperature at a given time. Thus, $B(t)$ will depend on core temperature $T_c(t)$, according to $\mathrm{d} B/\mathrm{d} t \sim -B/\tau_{ambip}$. The core temperature evolution is controlled by magnetic field decay heating and cooling due to neutrino and surface photon emission. We follow \citet{Lu2022} by taking the magnetic field heating rate $L_B \simeq V_{core} B^2 / (8 \pi \tau_{ambip}$) (where $V_{core} \simeq 4 \times 10^{18} \textrm{cm}^3$ is the volume of the outer core), photon cooling rate $L_{\gamma} \sim 9 \times 10^{33} T_{c,8.5}^{2.2} \textrm{erg} \ \textrm{s}^{-1}$ \citep{Gudmundsson1982, Page2004}, and the neutrino cooling rate $L_{\nu} \sim 5 \times 10^{35} T_{c,8.5}^{8} \ \textrm{erg} \ \textrm{s}^{-1} $ \citep{Shapiro1983}.

At early times and temperatures $T_c \gg 10^8$~K, the cooling will be dominated by neutrino emission. In this regime, if $L_{\nu} > L_B$, the temperature goes as $T_c \simeq 10^{8.5} (t/\textrm{kyr})^{-1/6}$~K. If magnetic decay heating dominates ($L_\nu < L_{\rm B}$), then the temperature will plateau at $T_c \simeq 2.8 \times 10^8 B_{15}^{0.4} L_5^{-0.2}$~K. At late time $t\geq t_{\rm tr} \sim 6 \times 10^4$~yrs (found when $L_{\nu} = L_{\gamma}$), if $L_{\gamma} > L_B$, then the temperature will go as $T_c \simeq T_{tr} (t/t_{tr})^{-5}$~K, where $T_{tr}\simeq 1.5 \times 10^8\rm\, K$ is the temperature at $t=t_{\rm tr}$. If heating dominates ($L_\gamma < L_{\rm B}$), then the temperature will plateau at $T_c \simeq 0.7 \times 10^8 B_{14}^{0.95} L_5^{-0.48}$~K. 

We numerically solve for $T_c(t)$ and $B(t)$. Using this and the inner disk accretion rate $\dot{M}_{in}(t)$ from \S\ref{sec:diskevo}, we calculate the torques on the NS and hence the spin period evolution for a given initial period $P_0$. We evolve the system until 1~Myr, beyond which the disk would either evaporate or become passive. We also note that our results do not depend strongly on the magnetic field decay prescription (see Appendix~\ref{sec:decay}). This is because most ULPs are produced within $10^3\rm\, yr$ after the NS formation (see Figure~\ref{fig:periodevo}), which is typically shorter than the magnetic field decay timescale.

\subsection{Spin Period Results}

We consider two initial spin periods of $P_0 = 0.033$~s and $P_0 = 0.1$~s. The initial spin period sets the light-cylinder radius $r_{lc}$, which in turn governs if the disk interacts with the NS. However, once the NS enters the propeller phase, the system will evolve to reach the equilibrium spin rate $\Omega_{eq}$ (independent of $P_0$, see Equation~\ref{eqn:omega_eq}), so our results are largely insensitive to the initial spin periods if the magnetic field is sufficiently large. 

\begin{figure}
    \centering
    \includegraphics[width=0.47\textwidth]{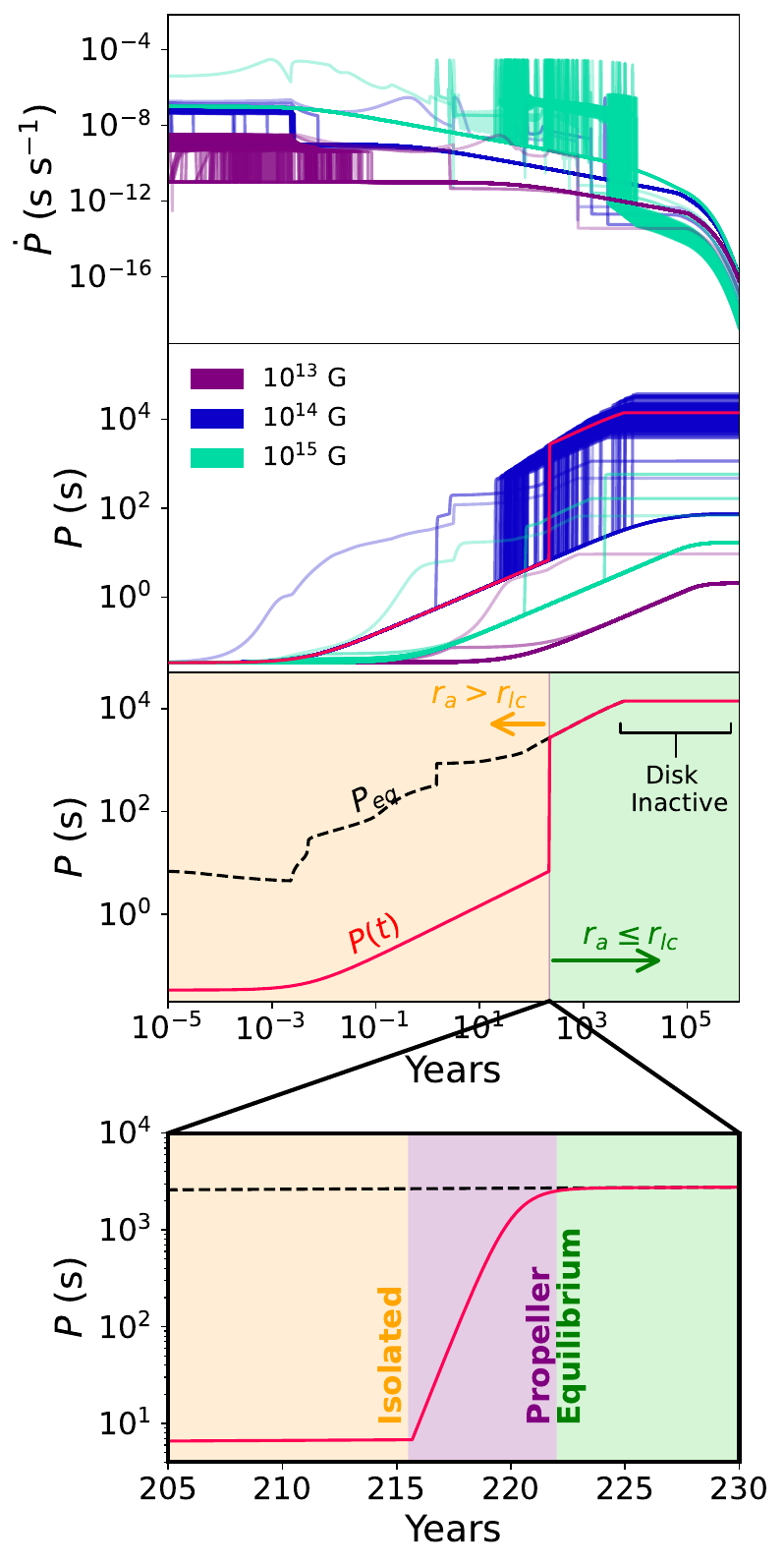}

    \caption{\textbf{Period evolution for all disks} formed from an initial kick velocity of $v_k = 400~\textrm{km} \ \textrm{s}^{-1}$, initial spin period $P_0 = 0.033$~s, and initial B-field strength of $10^{13}$~G, $10^{14}$~G, and $10^{15}$~G. Disks presented are formed from total mass capture. For disks undergoing the propeller phase, it appears they undergo a sudden jump in period. However, we zoom in on one of the systems (indicated in red), plotted with the equilibrium period calculated from Equation~\ref{eqn:omega_eq}. The sudden jump is physical and is from the nature of log-space. In reality, the NS reaches the equilibrium spin period over several years. Meanwhile, the early-time fluctuations in $\dot{P}$ for $B_0 = 10^{13}$~G is understood to be from $r_A$ being within $r_{lc}$, and then $\dot{P}$ suddenly drops as the disk viscously spreads and $r_A>r_{lc}$.}
    \label{fig:periodevo}
\end{figure}

Figure~\ref{fig:periodevo} shows the period evolution of all disks for NS kick velocity $v_k = 400~\textrm{km} \ \textrm{s}^{-1}$, initial spin period $P_0 = 0.033$~s, and initial B-field strengths $10^{13}$, $10^{14}$, and $10^{15}$~G. We find a bi-modal distribution of spin periods. The short-period mode is for NSs that never interact with the accretion disk and evolve as regular isolated pulsars.
The second mode is for NSs that undergo the propeller phase within the first $\sim$kyr of their lifetimes, resulting in a long spin period by 1~Myr. Our models show that the NS is in the propeller phase for a short amount of time, and rapidly spins down as it reaches the equilibrium period described by Equation~\ref{eqn:omega_eq}. We zoom in on one random disk in Figure~\ref{fig:periodevo} to show this short-term evolution. For this one example, the NS rapidly spins down around $t\simeq 200\rm\, yr$ and reaches a final period of $P\simeq 10^4\rm\, s$ at $t\gtrsim10\rm\, kyr$. 


\begin{figure*}
    \centering
    \includegraphics[width=\textwidth]{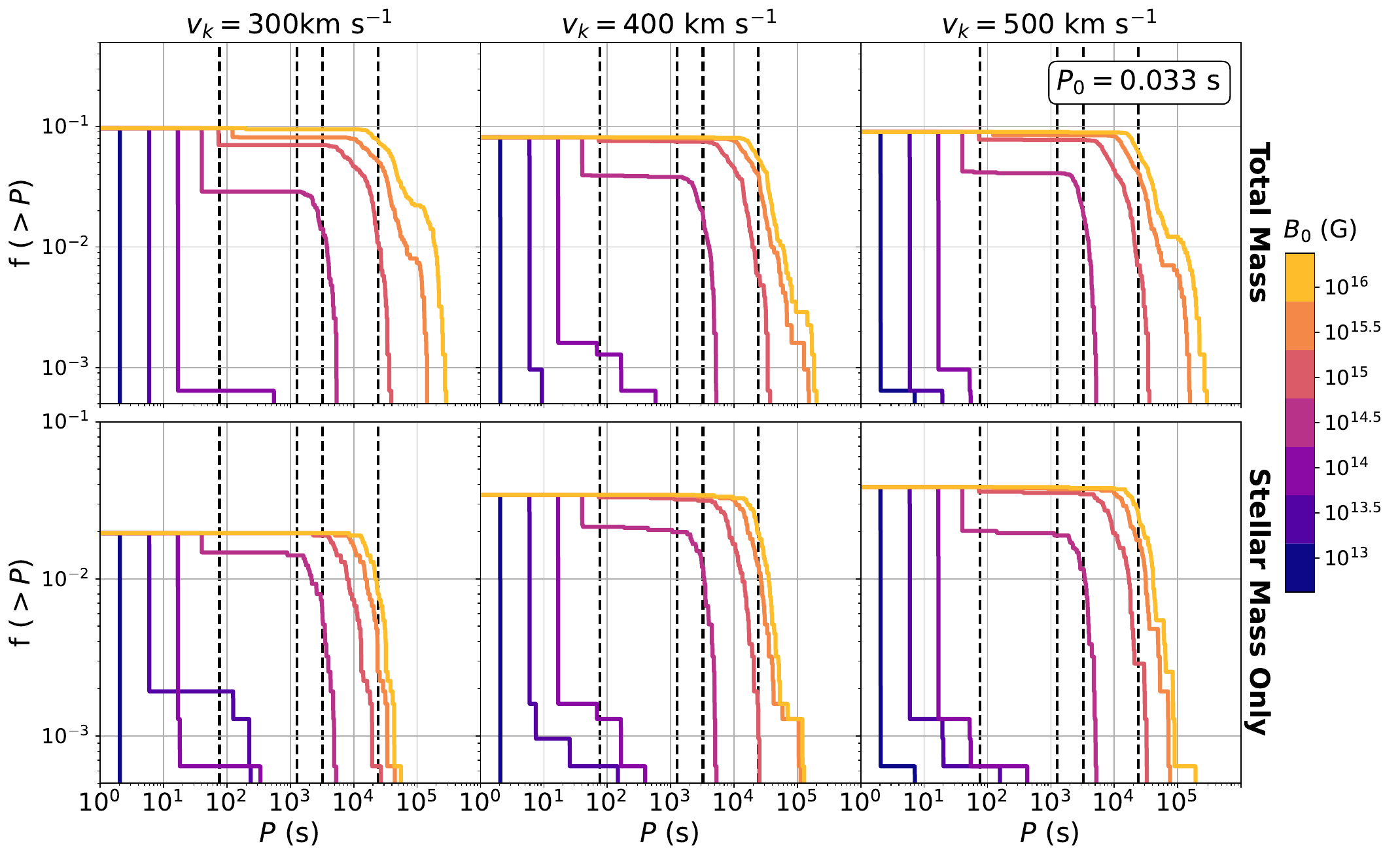}
    \caption{\textbf{Cumulative Distribution Plot of Spin Periods for  $P_0 = 0.033$~s.} Fractions are taken out of all possible kick angles. Columns are of different initial kick velocities, with the top row being for disks considering the total mass captured, and the bottom row being disks if we only considered stellar matter. Color indicates initial B-field strength. Dashed lines represent observed radio transients, PSR J0901-4046 at 76~s, GMP-1839 at 21~min, ASKAP-J1935 at 54~min, and IE-1613 at 6.7~hrs.}
    \label{fig:period_frac}
\end{figure*}

\begin{figure*}
    \centering
    \includegraphics[width=\textwidth]{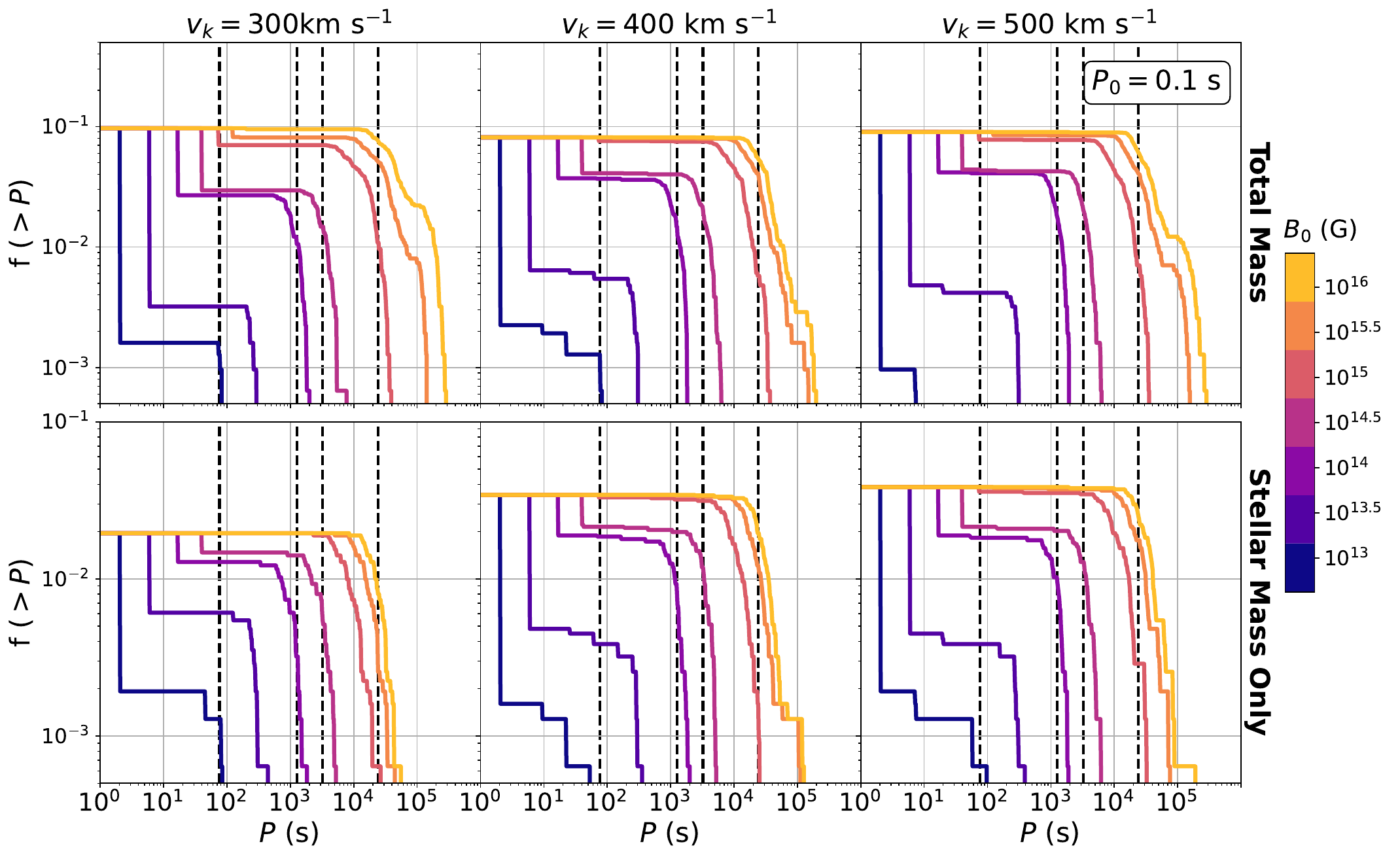}
    \caption{Same as Figure~\ref{fig:period_frac}, but for  $P_0 = 0.1$~s.}
    \label{fig:period_frac_20}
\end{figure*}

We study the spin-period evolution for the three initial kick velocities above, 
and tested seven different initial magnetic field strengths, ranging from $10^{13}$ to $10^{16}$~G in steps of 0.5 in log space. Figure~\ref{fig:period_frac} and Figure~\ref{fig:period_frac_20} show the cumulative distribution of spin periods by the end of our disk evolution at 1~Myr for the initial spin periods of 0.33~s and 0.10~s, respectively. 
We see the same bimodal distribution that is highlighted in Figure~\ref{fig:periodevo}: some pulsars never interact with an accretion disk and remain in the short-period mode, whereas others undergo the propeller phase during their evolution and end up in the long-period mode.

For stronger NS magnetic fields, we observe more ultra-long period pulsars. This is because a stronger magnetic field would extend the influence of the magnetosphere and increase the torque the disk provides on the NS. We do not observe any ultra-long period pulsars (defined as $P\geq10^3~$s for our purposes) for NSs with initial B-field strengths $\leq 10^{14}$~G when $P_0 = 0.033\rm\,s$, though a very small fraction do undergo the propeller phase. However, when we increase the initial spin period from $P_0=$ 0.033~s to $P_0=$ 0.1~s, the fraction of low magnetic field NSs that enter the propeller phase increases. It is possible to achieve pulsars with periods on the order of $\sim10^2\rm \, s$ with lower magnetic field strengths $\leq 10^{13.5}$~G and periods on the order $\sim10^3\rm \, s$ with magnetic field strengths $\sim 10^{14}$~G if we start with a higher initial period. For magnetic fields $\gtrsim 10^{14.5}$~G, our results are practically identical between the two initial spin periods.
 
\begin{table*}
\centering
\begin{tabular}{|c|c|c c c c c c c|} 
\hline 
& \textbf{$f(>P)$} & 13 & 13.5 & 14 & 14.5 & 15 & 15.5 & 16\\
\hline
$P_0 = 0.033~$s & $5\times10^{-2}$ & 2.0 & 5.9 & $1.7 \times 10^1$ &$4.0 \times 10^1$ & $9.2 \times 10^3$ & $2.7 \times 10^4$ & $4.4 \times 10^4$\\
 & $10^{-2}$ & 2.0 & 5.9 & $1.7 \times 10^1$ &  $4.1 \times 10^3$ & $2.5 \times 10^4$ & $6.8 \times 10^4$ & $1.9 \times 10^5$\\
& $5 \times10^{-3}$&2.0 & 5.9 & $1.7 \times 10^1$& $4.6 \times 10^3$ & $3.1 \times 10^4$ & $1.2 \times 10^5$ & $2.2 \times 10^5$\\
& $10^{-3}$ &2.0 & 9.4 & $1.7 \times 10^2$ & $5.3 \times 10^3$ & $3.7 \times 10^4$ & $1.5 \times 10^5$ & $2.7 \times 10^5$\\ 
& max P & 9.4 & $2.5 \times 10^1$& $5.8 \times 10^2$& $5.4 \times 10^3$& $3.9 \times 10^4$& $1.6 \times 10^5$& $2.9 \times 10^5$ \\
\hline
$P_0 = 0.1~$s & $5\times10^{-2}$ & 2.0 & 6.0 & $1.7 \times 10^1$ & $4.0 \times 10^1$ & $9.2 \times 10^3$ & $2.6 \times 10^3$  & $4.4 \times 10^3$\\
 & $10^{-2}$ & 2.0 & 6.0 & $1.5 \times 10^3$ & $4.4 \times 10^3$ & $2.5 \times 10^4$ & $6.8 \times 10^4$ & $1.9 \times 10^5$ \\
& $5 \times10^{-3}$ & 2.0 & $2.2 \times 10^2$ & $1.7 \times 10^3$ & $4.9 \times 10^3$ & $3.1 \times 10^4$ & $1.2 \times 10^5$ & $2.2 \times 10^5$\\
& $10^{-3}$ & $8.0 \times 10^1$ & $3.1 \times 10^2$ & $1.9 \times 10^3$ &  $6.2 \times 10^3$ &  $3.6 \times 10^4$ &  $1.5 \times 10^5$ &$2.7 \times 10^5$\\
& max P &  $8.3 \times 10^1$& $3.1 \times 10^2$& $2.0 \times 10^3$& $8.7 \times 10^3$& $3.9 \times 10^4$& $1.6 \times 10^5$& $2.9 \times 10^5$ \\
\hline 
\end{tabular}
\caption{Final spin period results from Figure~\ref{fig:period_frac} and Figure~\ref{fig:period_frac_20} for each given initial magnetic field strength (represented in log). For each fraction shown, the corresponding period is the minimum value achieved by that fraction of cases. Of the three velocities, the largest spin period at a given fraction is recorded. In the last row, we include the largest spin rates for each magnetic field strength. }
\label{table:cdf}
\end{table*}

In Table~\ref{table:cdf}, we report on the fractional occurrence rate of final spin periods for each magnetic field strength. We also include the largest spin period achieved in our simulations. Based on these values, we find that the initial period can affect the largest spin period achieved for lower initial magnetic fields ($B_0 = 10^{13}$~G,  $10^{13.5}$~G, and $10^{14}$~G). In fact, the spin periods for given cumulative occurrence fractions $f(>P)$ for $P_0 = 0.1$~s are about 10 times longer than those for $P_0 = 0.033$~s. However, for $B_0 \gtrsim 10^{14.5}$~G, the two initial periods do not lead to major differences in the final distribution for long-period pulsars (within runoff margins).

The period distribution of pulsars depends strongly on the initial magnetic field strength $B_0$. For example, using $P_0 = 0.033~\rm s$, we only observe periods up to $\sim500$~s (although rarely produced) for $B_0 = 10^{14}$~G. For a stronger initial field of $B_0 = 10^{14.5}$~G, the maximum NS period increase to roughly $5\times10^3\rm\, s$. Longer periods ($P\gtrsim 10^4\rm\, s$) are produced for $B_0\gtrsim10^{15}\rm\,G$. 

\begin{figure*}
    \centering
    \includegraphics[width=\textwidth]{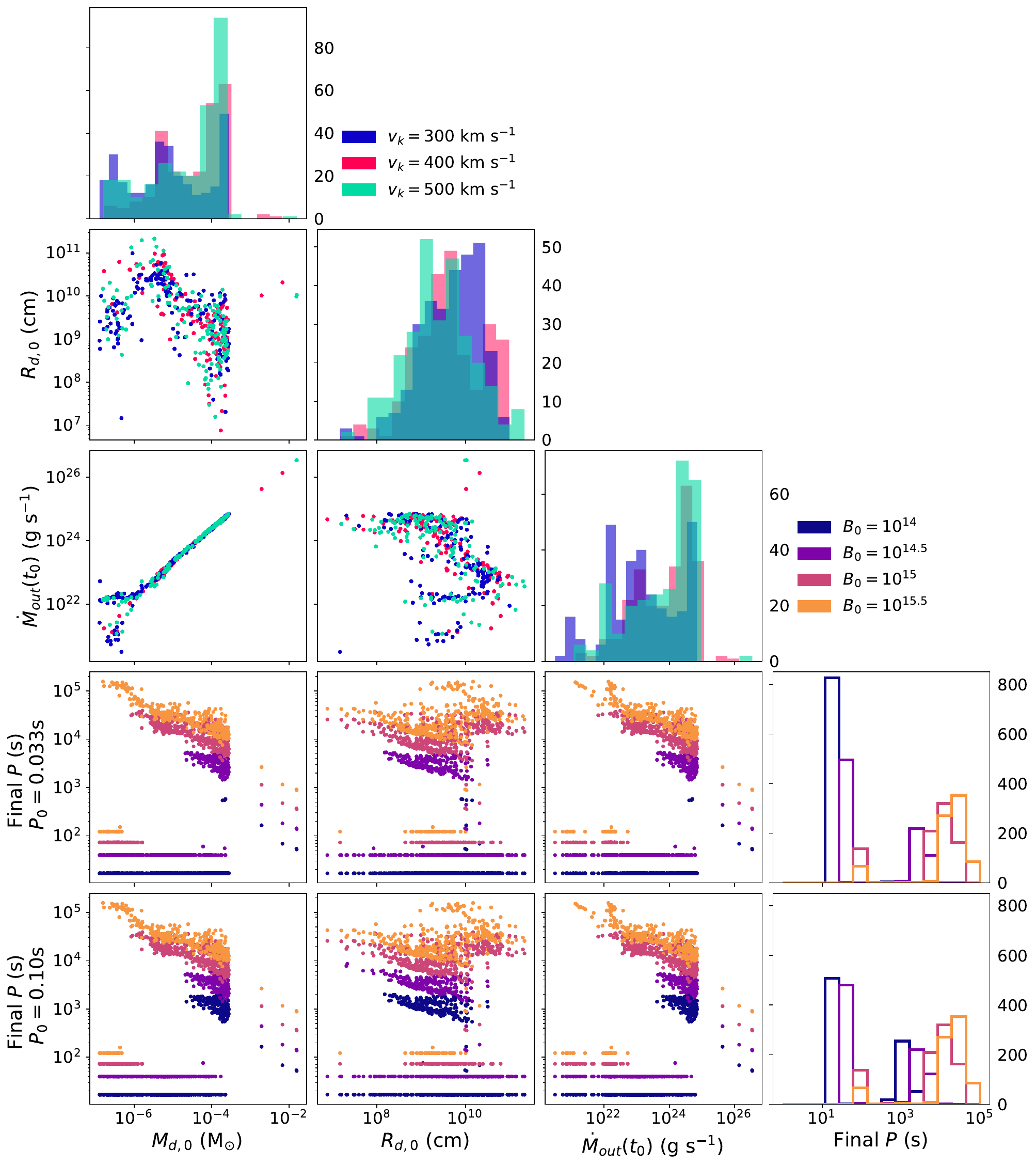}
    \caption{\textbf{Corner plot of initial disk parameters and final spin periods.} We separate the final spin period results by initial period. Colors of plots involving the spin periods indicate initial B-field strength and are for all kick velocities that we tested. Other plots not involving the final spin periods are separated by initial kick velocity. $M_{d,0}$ indicates the total captured mass,  $R_{d,0}$ is the initial circularization radius, and $\dot{M}_{\rm out}$ is the outer disk accretion rate at $t_0 =20$~hrs. }
    \label{fig:corner}
\end{figure*}

Figure~\ref{fig:corner} shows the final period's dependence on the disk properties, including the total captured mass, initial circularization radius, and disk accretion rate at $t = t_0=20\rm\, hr$ (chosen to be the end of disk assembly). We find that, for a given initial B-field strength, the final spin period is longer for \textit{lower} captured mass. This counter-intuitive result can be understood as follows. A higher captured mass makes it more likely for the system to enter the propeller phase and hence produce an ultra-long period pulsar. However, once the propeller phase is achieved, the NS quickly reaches the equilibrium spin period $P_{eq} = 2\pi/\Omega_{eq} \propto \dot{M}^{-3/7}B^{6/7}$ (Equation~\ref{eqn:omega_eq}), so the final period is longer for lower accretion rates. Lower accretion rates are achieved with smaller mass disks. 

\begin{figure*}
    \centering
    \includegraphics[width=\textwidth]{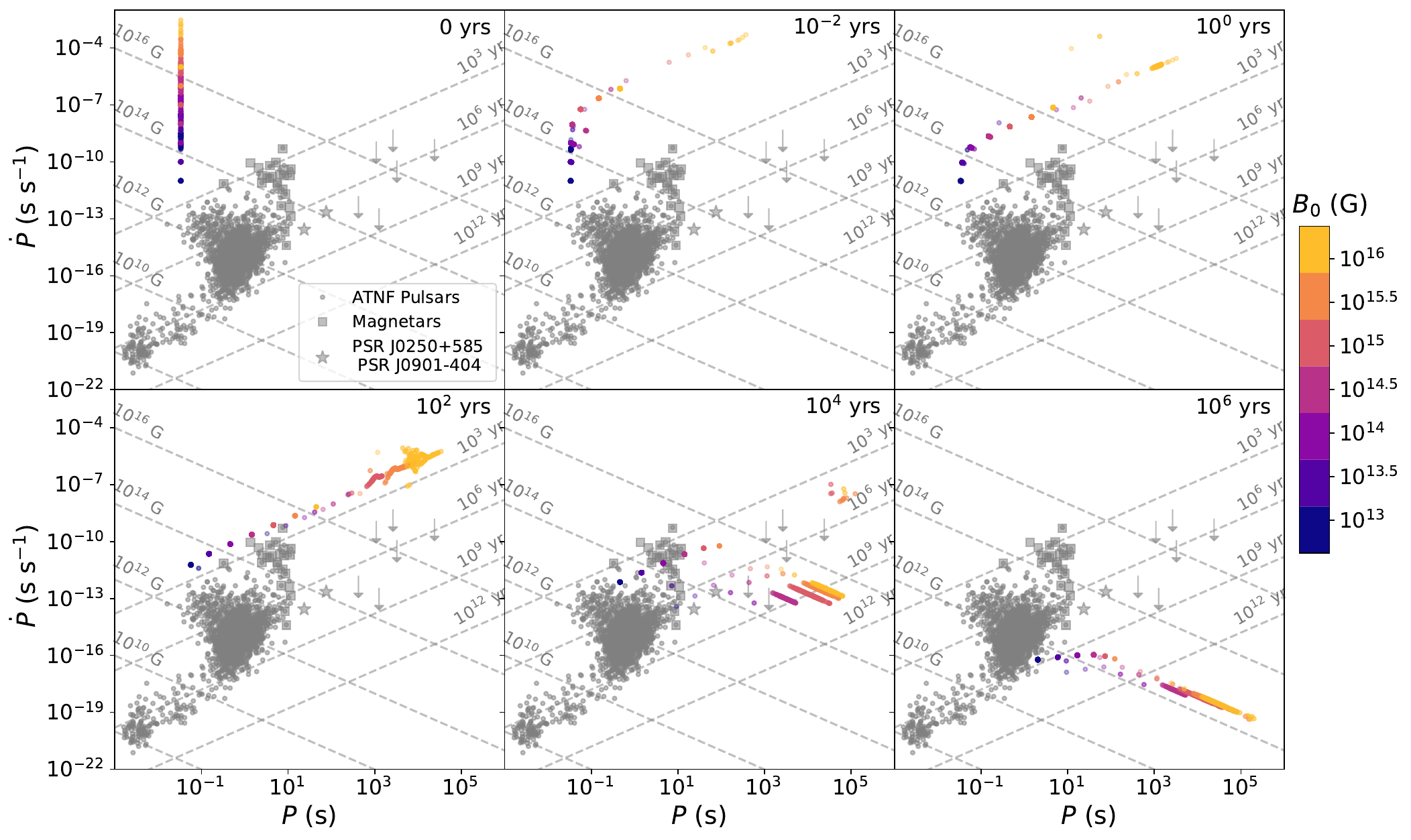}
    \caption{\textbf{$P-\dot{P}$ Diagram over time.} Colored points indicates initial B-field strength of our simulated NSs. Plotted additionally in gray are pulsars from the ATNF catalog \citep[][\url{http://www.atnf.csiro.au/research/ pulsar/psrcat}]{pulsarcat}, magnetars \citep[][\url{http://www.physics.mcgill.ca/~pulsar/magnetar/main.html}]{magcat}, and long period pulsars PSR J0250 and PSR J0901 \citep{tan2018,caleb2022}. Other recently discovered long period radio transients are indicated by downwards arrows to show their upper limits \citep{chime421,18min,21min,54min,DeLuca2006}. Lines of equal B-field strength are calculated from spin-down due to magnetic dipole radiation: $\dot{P} = (4 \pi^2 B)/ (3Ic^3P)$. Lines of equal age are calculated using the spin-down assuming magnetic braking: $\dot{P} = P/(2\tau)$, where $\tau$ is the characteristic age. Although this figure is for NSs with $P_0 = 0.033$~s, for $B_0 \geq 10^{14}$~G, our results do not significantly change. We keep our plot within reasonable margins of what has been observed by current pulsars. }
    \label{fig:PPdot}
\end{figure*}

We note that our spin rates are calculated at 1~Myr and that not all ULPs are assumed to be at this age. However, we find that the accretion disks slow down the NSs within the first $10\,$kyr (see Figure~\ref{fig:periodevo}). For ages $\geq 10$~kyr, the only torque is due to magnetic dipole emission, which does not affect the long-term evolution of pulsars if the period is already very long. With this in mind, we look at the $P - \dot{P}$ evolution over time for these pulsars, as plotted by Figure~\ref{fig:PPdot}. For the strongest initial magnetic fields, we observe NSs with spin periods $10^2 \sim 10^3$~s within one year. By 100 years, a larger fraction of the NS population (particularly those with initial B-fields $\geq 10^{14}$~G) reaches spin periods $ > 10^2 $~s. 

The NSs in our model experience high $\dot{P}$ values within their first year of evolution. This can be understood to be due to a combination of magnetic dipole emission and the NSs entering the propeller phase, where they undergo rapid spin-down as their spin periods reach their equilibrium periods. After $\gtrsim 10^4$ years, $\dot{P}$ is very low as it is only caused by magnetic dipole emission, which provides little torque on the NSs (given the long period and weakened field strength).

\section{Discussion} \label{sec:discussion}

In this section, we compare our results to observations, and then we discuss the limitations of our model as well as the broader implications that are not studied in detail in this work.

\subsection{Comparison to Observations}

Current observations of pulsars and ULPs suggest that there might be a bimodal distribution in the period distribution --- canonical pulsars have spin periods $\leq12$~s and ULPs have periods of minutes to hours. However, we note that no statistical study has been done to definitely show the bimodal distribution predicted in this work. 
These two distinct groups of periods may be due to observational biases, and in the upcoming years there may be more radio pulsators discovered with intermediate spin periods. In our model (and in fact all ULP models based on NSs spinning down due to a propeller phase), the period distribution is bimodal for the following reason. A NS either never interacts with its accretion disk (in case of weak initial magnetic fields and low disk masses) and evolves as a canonical NS with a fast spin, or once the NS magnetosphere interacts with the accretion disk, the system enters the propeller phase where the NS is rapidly spun-down to very long periods of $10^2\rm\, s$ up to $10^5\rm\, s$. We encourage a future statistical study to demonstrate or rule out such a bimodal distribution of spin periods in the current sample of pulsars and ULPs.

In terms of individual spin periods, our model predicts short ULPs, such as PSR J0901-4046 with $P=76\rm\, s$ \citep{caleb2022}, which are produced even for relatively low initial magnetic field strengths $10^{13}\rm\, G$, as long as the initial periods are sufficiently long $P_0\gtrsim 0.1\rm \,s$ (such that the system enters the propeller phase). Within our model, such short ULPs are produced for initial field strengths up to $10^{14.5}\rm\, G$. On the other hand, the longest ULPs, such as IE-1613 with $P=6.7\rm\, hrs$ \citep{DeLuca2006}, are only produced for relatively strong initial field strengths of $B_0\gtrsim10^{15}\rm\,G$ --- consistent with typical magnetars.

Our model also can explain the upper limits of period derivative for the observed ULPs. During the propeller phase, the period derivative becomes extremely large $\dot{P}$ up to $10^{-5}$, which is easily measurable. However, since the propeller phase is extremely short-lived (lasting for decades), it is difficult to catch the evolution observationally. Once the NS reaches the equilibrium spin, then the period changes on the disk evolutionary timescales of 1 to 10 kyrs, which gives rise to $\dot{P}\sim 10^{-9}$ to $10^{-7}$ ($10^3\rm\, s$ to $10^5\rm\, s$ divided by $10\rm\, kyr$). However, during the equilibrium-spin phase, the signatures of the accretion disk may still be detectable, e.g., in the form of mid-infrared emission from the X-ray-heated disk. The infrared emission from the outer disk has luminosity (Equation~\ref{eqn:irradiation}) and effective temperature
\begin{equation}\label{eqn:IR_emission}
\begin{split}
    L_{\rm IR} &\simeq 3\times 10^{32}\mathrm{\,erg\,s^{-1}}\, \frac{L_x}{ 10^{35}\rm\, erg\,s^{-1}} \frac{H/R_d}{ 0.03}, \\
    T_{\rm IR} &\simeq 250\mathrm{\,K}\, \left(\frac{L_x}{ 10^{35}\rm\, erg\,s^{-1}} \frac{H/R_d}{0.03}\right)^{1/4} \left(\frac{R_{d}}{\rm AU}\right)^{-1/2},
\end{split}
\end{equation}
where $L_x$ is the X-ray luminosity of the neutron star, $H/R_d\sim 3\%$ is the typical aspect ratio in our model, and $R_d$ is the outer disk radius.
More observational and theoretical works along this direction are needed to see if the disk signatures are observable. On longer timescales $t\gtrsim10\rm\,kyrs$ (which is the case for most observed systems), the disks either evaporate or become inactive. In this case, the NS spin-down is only due to magnetic dipole emission, which produces a very low period-derivative $\dot{P}\sim 10^{-18} (P/10^3\mathrm{\,s})^{-1} (B/10^{12}\mathrm{\,G})^2$, which are not detectable for ULPs. In our model, the observed ULPs are consistent with relatively old systems where the disks have become inactive. Future observations may detect the more rare cases in the equilibrium-spin phase with much larger $\dot{P}$.

We additionally discuss the formation rate of ULPs. In the following, we calculate our rates under the assumption that ULPs require magnetar-level initial field strengths $B_0\gtrsim10^{14}\rm\, G$ (although some short ULPs may be produced with $B_0$ as low as $10^{13}\rm\, G$).

We start from the core-collapse rate of 3 per century in our Galaxy \citep[][and cites therein]{KeaneKramer2008} and assume that a fraction $f_{\rm mag} \sim 10\%$ of NSs are formed as magnetars \citep[although this fraction could be larger, e.g.,][]{Beniamini2019, GillHeyl2007, KeaneKramer2008}. Based on the rate of type-Ib/c supernovae \citep[e.g.,][]{Shivvers:2017aa}, we assume that $30\%$ of NSs are born in a close binary as studied in this work. If magnetar formation are not correlated with stripped-envelope supernovae, we obtain a formation rate of magnetars born in stripped-envelope systems to be of the order $10^{-3}\rm\, yr^{-1}$ in our Galaxy.

As shown in Figs. \ref{fig:period_frac} and \ref{fig:period_frac_20}, we find the fractional outcome of ULPs with $P\gtrsim 10^3\rm\, s$ to be $f_{\rm ULP}\sim 3\%$ to $10\%$, considering different kick velocities, initial magnetic field strengths, and initial NS spin periods. In this work, we only consider a pre-supernova binary separation of $20 ~\rm R_\odot$, which corresponds to orbital period $P_{\rm orb}\simeq 3\rm\, d$.
We expect the occurrence fraction of ULPs to increase for shorter orbital periods for various reasons. A larger fraction of the ejecta would be intercepted by the companion star, the companion star is more strongly modified by a larger ram pressure, and a shorter orbital separation also increases the chance of closer encounters between the NS and companion star. Thus, we estimate the ULP formation rate to be
\begin{equation}\label{eq:ULP_rate}
    \dot{N}_{\rm ULP} \sim 10^{-4}\rm yr^{-1}\, \frac{f_{\rm ULP}}{10\%}\, \frac{f_{\rm mag}}{10\%},
\end{equation}
where $f_{\rm ULP}$ is the fractional outcomes of ULPs from magnetars born in close binary systems and $f_{\rm mag}$ is the magnetar fraction.
If all NSs are able to emit radio emission for up to $t_{\rm life}\sim 1\rm\, Myr$, then the total number of radio-bright ULPs in the Milky Way is given by $N_{\rm ULP}\sim 10^2$ for our fiducial values in eq. (\ref{eq:ULP_rate}).

Note that our estimate suffers from a number of major uncertainties that are beyond the scope of this work: (1) the magnetar formation rate may be higher than $f_{\rm mag}\sim 10\%$ \citep[][whose population study prefers $f_{\rm mag}\sim 40\%$]{Beniamini2019}; (2) if stripped-envelope supernovae preferentially make magnetars \citep[due to e.g., faster core rotation from tidal spin-up in binary interactions,][]{Fuller:2022aa}, one may effectively take $f_{\rm mag}\sim 100\%$ within the stripped-envelope supernova population; and (3) the radio emission lifetime for ULPs (arbitrarily taken to be 1~Myr in the above estimate) is unknown, although the lack of bright persistent X-ray emission from the observed ULPs may suggest that they are likely much older than the known Galactic magnetar population whose typical ages are 1--10~kyrs. If we factor in all these uncertainties (while keeping $t_{\rm life} \sim 1\rm\, Myr$), we find $10\lesssim N_{\rm ULP}\lesssim 10^3$. In the most optimistic case of $N_{\rm ULP}\sim 10^3$, the nearest ULP may be as close as $300\rm\, pc$ away from the solar system.

Despite the uncertainties, we conclude that our model should contribute a non-negligible fraction of observable ULPs in the Milky Way.

\subsection{Limitations}

In this subsection, we discuss the limitations of our model, which may be improved in future works.

First, our model can be improved by considering a radially resolved disk (instead of a one-zone model) and a more physical consideration for evaporation. This would allow us to capture the radial dependence of irradiation and evaporation --- this is important to capture the final fate of the NS and the disk. 


Second, our mass capture estimates are also limited by the set-up of the hydrodynamic simulation. We do not consider the realistic mass capture by the gravitational potential of a NS moving in an unbound trajectory, as it is more computational heavy to do so. Moreover, the captured slow-moving gas near the center of explosion is considered as a part of ``fallback accretion'' and hence removed from our disk modeling (as described in \S~\ref{sec:masscapture}). This is because the gas near the explosion center is likely not properly modeled by our $\gamma = 5/3$ equation of state and our initial conditions are also likely simplistic. This makes our mass capture estimates conservative --- it is possible that fallback accretion also contributes to disk formation and that a larger fraction of NSs evolve into ULPs.

Third, we do not survey a broad parameter space in companion star masses and orbital separations. We expect our results, especially the fractional outcome of ULPs, to depend strongly on the orbital separation. This is mainly because (1) the ram pressure $p_{\rm ram}$ of the ejecta at the position of the companion star scales with the orbital separation $a_{\rm sep}$  as $p_{\rm ram}\propto a_{\rm sep}^{-3}$ and (2) the solid angle $\Omega_*$ spanned by the companion star (of a given radius) scales with the orbital separation as $\Omega_*\propto a_{\rm sep}^{-2}$. For these reasons, we expect the fractional outcome of ULPs to increase for shorter orbital separations. Unfortunately, the pre-supernova orbital separations of stripped-envelope supernovae are currently highly uncertain.

\subsection{Other Consequences}


In this subsection, we briefly mention other consequences of our model that are not directly related to ULPs.

Although in this paper we only consider unbound orbits, observations and binary population synthesis models suggest that about 5\% (or higher) of the time a neutron star will remain bound to its companion post core-collapse \citep[e.g.,][]{Chrimes2022C}. If the NS remains bound to the companion, it will repetitively pass through the companion’s inflated envelope (see Figure~\ref{fig:athena_ex}) and hence produce periodic power injections into the supernova ejecta as suggested by \citet{2022MNRAS.517.4544H}. This can potentially explain a recently discovered stripped-envelope supernova, SN 2022jli \citep{Chen2024}. SN 2022jli showed a 12.4-day periodicity in its brightness variations and H$\alpha$ velocities \citep{Chen2024}; this periodicity is likely caused by the remnant compact object being in a bound orbit with its companion post supernova. SN 2015ap also showed possible modulations in its light curve with a 8.4~day periodicity that could similarly be described by binary interactions \citep{2025arXiv250315977R}. 




If magnetars form in close binaries, our work shows that a fraction ($\sim 10\%$, depending on the orbital period) of them are born with accretion disks. An accretion disk could explain the excess infrared emission from magnetars, such as 4U 0142+61 and 1E 2259+586 \citep{Hulleman2004, Wang2006, Ertan2007,Kaplan2009}. As more infrared observations are made on magnetars \citep[e.g.,][]{Hare2024}, we may be able to constrain the percentage of magnetars that have accretion disks and compare it to our predictions. Furthermore, there have been multiple planets found orbiting pulsars \citep[e.g.,][]{Wolszczan1992}. Our work can be used to determine if the disk contains sufficient mass and angular momentum to form such exoplanets.

\section{Summary}
\label{sec:conclusion}

In this paper, we propose that a neutron star born in a close binary system can acquire an accretion disk by interacting with the companion star's shock-inflated envelope. We focus on the cases where the neutron star becomes unbound from the binary due to a natal kick. We follow the long-term disk evolution and its interaction with the neutron star's magnetosphere. These neutron stars may be observed as isolated pulsars whose spin periods are affected by the additional torque from the accretion disk. Our results are summarized as follows:

\begin{enumerate}
\item At a pre-supernova binary separation of $20 \rm ~R_\odot$ (or orbital period 3 days), we find that the neutron star becomes isolated with a surrounding accretion disk 8-10\% of the time. We expect the disk formation probability to increase towards closer binary separations.

\item The disks formed in our simulation have captured masses in the range $10^{-7} \sim 10^{-2} \ \mathrm{M}_{\odot}$, initial angular momentum of $10^{44} \sim 10^{49}$ erg~s, and initial circularization radii of $10^6 \sim 10^{11}$~cm. Compared to possible disk formation from supernova fallback, our binary model provides a new channel for disk formation with higher angular momenta.

\item We find that long-term interactions between the accretion disk and the neutron star's magnetosphere lead to a bimodal distribution of spin periods. The majority of neutron stars born in close binaries belong to the short-period mode with $P\lesssim 12\rm\, s$ because their spins have not been affected by the accretion disk. A minority (roughly $1\%$ to $10\%$) of neutron stars born in close binaries evolve to the long-period mode with $P\gtrsim 10^3\rm\, s$ as a result of a short-lived propeller phase which occurs almost exclusively for initial magnetic field strengths of $B_0\gtrsim 10^{14}\rm\, G$ --- typical for magnetars in the Milky Way. The long-period mode is consistent with observed ULPs such as CHIME J0630+25, GLEAM-J1627, ASKAP-J1935, and IE 161348-505, provided that these objects initially had magnetar-level field strengths and underwent a propeller phase in their evolution. A small fraction ($0.1\%$ to $1\%$) of pulsars with moderate $10^{13}\lesssim B_0\lesssim 10^{14}\rm\, G$ and relatively slow initial periods $P_0\gtrsim 0.1\rm\, s$ evolve to $P\sim 10^2\rm\,s$ (similar to PSR J0901-4046) and fill in the gap between the two period modes.

\item We find that longer periods are produced for stronger initial field strengths: $P \sim 10^4\rm\, s$ requires $B_0\sim 10^{14.5}\rm\, G$, $P\sim 10^5\rm\, s$ requires $B_0\sim10^{15}\rm G$, and only magnetars with $B_0\gtrsim 10^{15}\rm\, G$ evolve to the longest periods $P\gtrsim 10^5\rm\, s$. This is because, soon after the system enters the propeller phase, the neutron star reaches the equilibrium spin $P_{\rm eq}\propto B^{6/7}$ (Equation~ \ref{eqn:omega_eq}). 

\item We predict a ULP formation rate of $\dot{N}_{\rm ULP} \sim 10^{-4}\rm\,yr$ for our fiducial parameters (eq. \ref{eq:ULP_rate}), but a broad range of $10^{-5} \lesssim \dot{N}_{\rm ULP}\lesssim 10^{-3}\rm\, yr^{-1}$ is allowed. Thus, the total number of ULPs with $P\gtrsim10^3\rm\,s$ in the Milky Way is in the range of $10\lesssim N_{\rm ULP}\lesssim 10^3$ for an active lifetime of 1 Myr. The main uncertainties in our estimate are (1) the magnetar fraction of neutron stars born in close binaries ($10\%\lesssim f_{\rm mag}\lesssim 100\%$), and (2) the occurrence fraction of disks ($f_{\rm disk}$). We find that $f_{\rm disk}\sim 10\%$ for a pre-supernova separation of $20 \rm ~R_\odot$, but it depends sensitively on separation. 

\item On a timescale of 10 kyrs, once the accretion disk evaporates (due to X-ray irradiation) or becomes inactive (due to mid-plane temperature $T\lesssim10^2\rm\, K$), we predict that the period derivatives of ULPs drops to very small values $\dot{P}\lesssim 10^{-18}$ that are not detectable by current observations. However, there is a small population of magnetars younger than 10 kyrs with active disk interactions, and in these cases, we predict detectable period derivatives of $10^{-9}\lesssim \dot{P} \lesssim 10^{-7}$, because the neutron star is kept at the equilibrium spin rate as the disk evolves.

\item The X-ray irradiation of the disk produces mid-infrared emission with $L_{\rm IR}\sim 10^{32}\rm\, erg\,s^{-1}$ and typical effective temperature $T_{\rm IR}\sim 250\rm\, K$ for the outer disk. The mid-infrared emission only lasts for a relatively short duration of $\sim\!10$ kyr.

\item We also predict that, for bound post-supernova orbits, the neutron star will periodically pass through the shock-inflated envelope of the companion star and that the accretion power may produce brightened, periodically modulated emission following a type-Ib/c supernova. This model may potentially explain the peculiar properties of SN2023jli \citep{Chen2024}.

\end{enumerate}

\begin{acknowledgments}
We thank Bruce Grossan, J. J. Eldridge, and Ping Chen for useful conversations. The research of S.C. and W.L. was supported by the Hellman Fellows Fund. This research benefited from interactions at workshops funded by the Gordon and Betty Moore
Foundation through grant GBMF5076 and through interactions at the Kavli Institute for Theoretical Physics, supported by NSF PHY-2309135. 
C. L. acknowledges support from the Miller Institute for Basic Research at UC Berkeley.
T.L.S.W. acknowledges support from the Gordon and Betty Moore Foundation through grant GBMF5076.
This research used the Savio computational cluster resource provided by the Berkeley Research Computing program at the University of California, Berkeley (supported by the UC Berkeley Chancellor, Vice Chancellor for Research, and Chief Information Officer). 
This research used resources of the National Energy Research Scientific Computing Center (NERSC), a Department of Energy User Facility (project m2218-2025).
\end{acknowledgments}

\software{\textsc{athena++} \citep{athena}, \texttt{numpy} \citep{2020Natur.585..357H}, \texttt{scipy} \citep{2020SciPy-NMeth}, \texttt{matplotlib} \citep{Hunter:2007}}


\appendix
\section{Resolution Tests} 
\label{sec:resolution}
\begin{figure}[h!]

    \centering
    \includegraphics[width=0.7\textwidth]{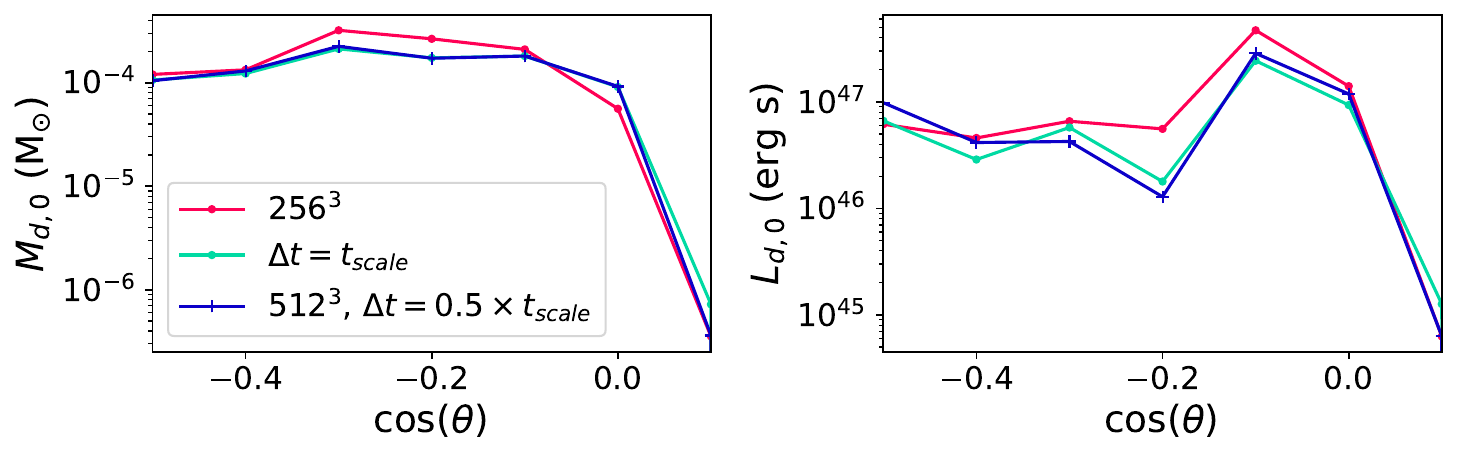}
\caption{Time and spatial resolution tests. We consider kicks in the cos$(\varphi_k)=1.0$ plane for $v_k = 400~\rm km~s^{-1}$ as an example, and show the total mass and angular momentum captures for each angle $\theta = \pi/2 - \theta_k$. Angles not included either are bound orbit trajectories or resulted in negligible mass capture.}
    \label{fig:res}
\end{figure}

In this section, we describe our convergence tests on the spatial and time resolutions of our simulations. 

First, we run another simulation with the same parameters as described in \S\ref{sec:hydro}, except we decrease the spatial resolution to a grid size of $256^3$ to compare to our original $512^3$ grid. Using the same Bondi capture methods described in \S\ref{sec:masscapture}, we compare the captured mass and angular momentum between the two grid sizes. 

Secondly, for our original simulation on a $512^3$ grid, we decrease the time in between each 3D snapshot data dumb to test the time resolution of our Bondi mass capture method. As described in \S\ref{sec:masscapture}, our fiducial 3D snapshot dump is once every $\delta t = 0.5 \times t_{scale}$. We double the time interval between each snapshot dump and run our Bondi mass capture as described in \S\ref{sec:masscapture}. 

The convergence of our results are demonstrated in Figure~\ref{fig:res}.

\section{Disk X-Ray Heating} \label{sec:diskgeometry}

\begin{figure}[h!]
    \centering
    \includegraphics[width=0.5\textwidth]{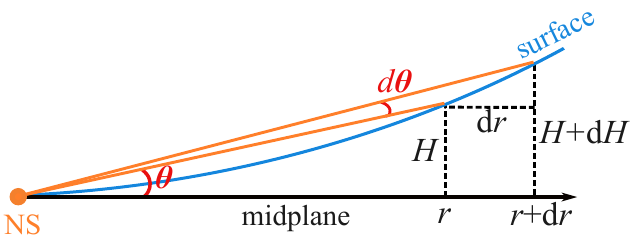}
    \caption{Geometry for disk heating in the case of a point-source.}
    \label{fig:disksurface}
\end{figure}

In this section, we derive the incident flux on the surface of a flared disk by treating the NS as a point source. We assume that the scale height of the disk scales with radius as a power-law $H \propto r^n$, with $n = \textrm{dln}H/\textrm{dln}r$. Let $\theta$ be the angle of the disk surface from the mid-plane (see Figure~\ref{fig:disksurface}), with $\textrm{tan}\theta = H/r$. The solid angle spanned by the annulus from $r$ to $r+\mathrm{d}r$ is $\textrm{d}\Omega = 2\pi \textrm{cos}\theta \textrm{d}\theta$, and
\begin{gather}
    \textrm{d}\theta = \textrm{cos}^2\theta \textrm{d}(H/r) = \textrm{cos}^2 \theta \frac{H}{r^2}(n-1)\textrm{d}r . 
\end{gather}
The corresponding differential disk surface area is
\begin{equation}
\textrm{d}A = 2\pi r \sqrt{(\textrm{d}H)^2 + (\textrm{d}r)^2} = 2\pi r \textrm{d}r\sqrt{1 + n^2 \textrm{tan}^2\theta}. 
\end{equation}
Thus, the flux per unit surface area of the disk is
\begin{equation}
F(r) = \frac{L \textrm{d}\Omega /4\pi}{\textrm{d}A} = \frac{L}{4 \pi r^2} {\frac{H}{r}} \frac{\textrm{cos}^3\theta (n-1)}{\sqrt{1+n^2\textrm{tan}^2\theta}}.
\end{equation}
At sufficiently late time ($t\gtrsim 1\rm\, $week), we find $\theta \sim H/r \ll 1$
and hence
\begin{gather}
    F(r) \approx \frac{L}{4 \pi r^2}\frac{H}{r}(n-1),
\end{gather}

Depending on the regime the disk is in, there are several solutions to $n$. For example, \citet{ChiangGoldreich1997} finds a value of 9/7 for a passive disk, whereas \citet{Frank2002} provided $n = 9/8$ for an accreting disk with viscous dissipation under the Kramer's opacity scaling law. We taken an intermediate value of $n=6/5$ in this work.

\section{Viscosity Dependency} \label{sec:alpha}

\begin{figure}[h!]
    \centering
    \includegraphics[width=\textwidth]{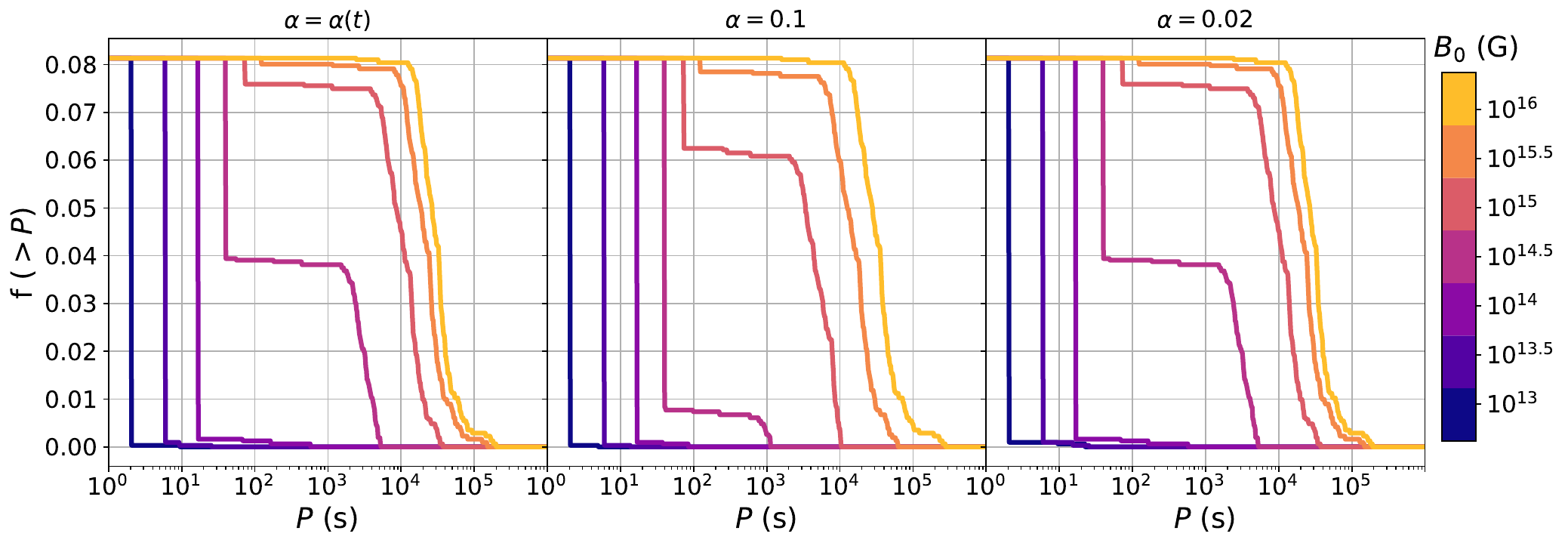}
    \caption{\textbf{Final period distributions for different $\alpha$ models}. We fix $v_k = 400\, \textrm{km} \, \textrm{s}^{-1}$ and $P_0=0.033$~s. $\alpha (t)$ corresponds to the default model in Equation~(\ref{eqn:alpha}).}
    \label{fig:alpha}
\end{figure}

Our disk evolution model adopts a varying viscousity parameter $\alpha$ that transitions from the hot regime $\alpha_{\rm hot}=0.1$ to the cold regime $\alpha_{\rm cold}=0.02$. In Figure~\ref{fig:alpha}, we show the final period distribution for constant $\alpha=0.02$ and $\alpha=0.1$ (all else are kept the same). We find that the results in our default model are very similar to the case with a constant $\alpha=0.02$. As for the constant $\alpha=0.1$ case, the results are still qualitatively similar in that ULPs can still form but with a slightly lower rate.

\section{B-Field Decay Dependency} \label{sec:decay}

\begin{figure}[h!]
    \centering
    \includegraphics[width=\textwidth]{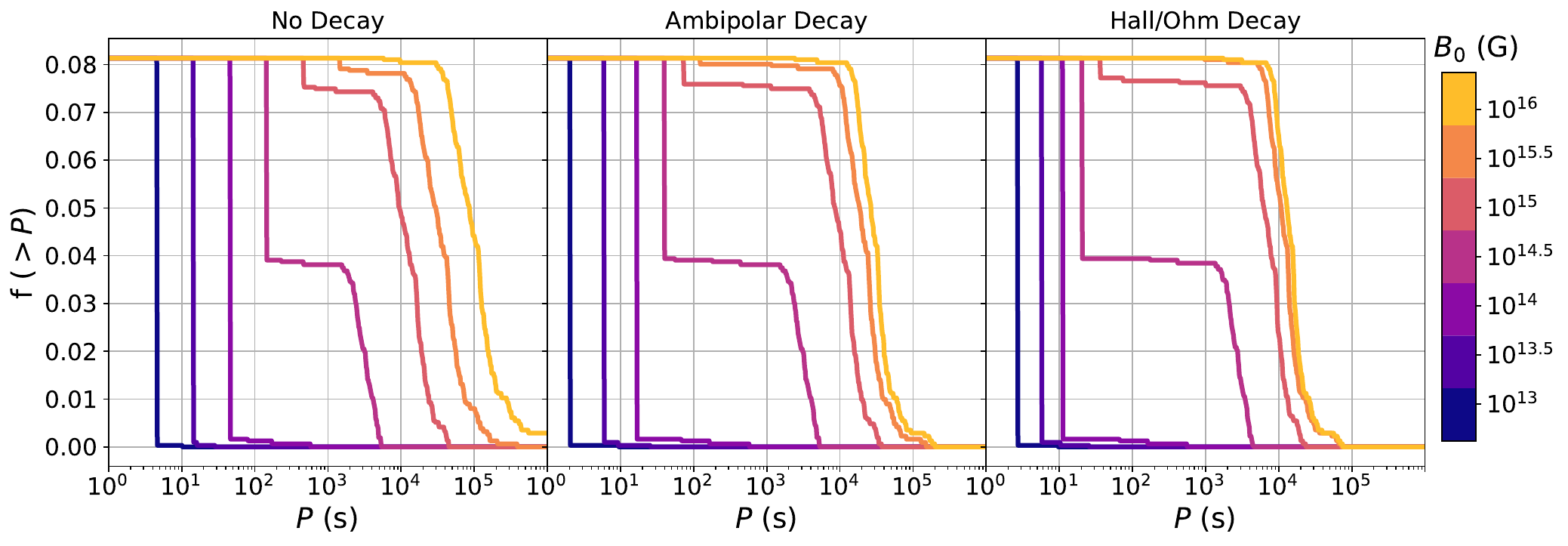}
    \caption{\textbf{Final period distributions for different field decay models}. We fix $v_k = 400\, \textrm{km} \, \textrm{s}^{-1}$ and $P_0=0.033$~s.}
    \label{fig:decay}
\end{figure}

We also considered a different field decay model as described in \citet{aguilera08_Bfield_decay}. In their model, the magnetic field evolves according to timescales of $\tau_{\textrm{Hall}} = 4\pi e n_e L^2 / c B_0$ for the Hall effect, and $\tau_{\textrm{Ohm}} = 4 \pi \sigma L^2 / c^2$ for ohmic dissipation, where $e$ is the election charge, $n_e$ is the average electron density in the crust, $\sigma$ is the conductivity, and $L$ is the scale length for the crust. Using the prescription in \citet{Ronchi2022}, we take $n_e = 10^{35} \ \textrm{cm}^{-3}$, $L = 1$~km, and $\sigma = 10^{24} \ \textrm{s}^{-1}$, leaving us with $\tau_{\textrm{Hall}} \sim 6.4 \times 10^4 \ \textrm{yr} \ B_{0,14}^{-1}$ and $\tau_{\textrm{Ohm}} \sim 4.4 \times 10^6 \ \textrm{yr}$ (with $B_{0,14} = B_0 / 10^{14}~$G). The magnetic field decay is dominated by $\tau_{\textrm{Hall}}$ at earlier times and by $\tau_{\textrm{Ohm}}$ at later times, as follows:
\begin{equation}
    B(t) = B_0 \frac{e^{-t/\tau_{\textrm{Ohm}}}}{1 + \frac{\tau_{\textrm{Ohm}}}{\tau_{\textrm{Hall}}}\left[ 1 - e^{-t/\textrm{Ohm}}\right]}. 
\end{equation}
The final spin period distributions for different field decay prescriptions are shown in Figure~\ref{fig:decay}. The similarity of the results means that our conclusions are not very sensitive to the field decay models considered in the literature, and this is mainly because the propeller phase usually occurs within the first 1 kyrs of evolution before the magnetic fields decay substantially. 

\bibliography{ULP.bib}{}
\bibliographystyle{aasjournalv7}



\end{document}